\documentclass[12pt]{iopart}

\expandafter\let\csname equation*\endcsname\relax
\expandafter\let\csname endequation*\endcsname\relax
\usepackage{amsmath}
\usepackage{iopams}
\usepackage{graphicx}
\usepackage{bm}
\usepackage{microtype}
\usepackage{placeins}
\usepackage{cite}
\begin{document}

\title[TTN-RC: Hierarchical Ensemble]{Tree Tensor Network Reservoir Computing: \\
Hierarchical Ensemble with Invariant Phase Boundaries}
\author{Daiki Sasaki$^{1}$, Chih-Chieh Chen$^{2}$, and Tomah Sogabe$^{3}$}
\address{$^{1}$Department of Engineering Science, The University of Electro-Communications, Chofu, Tokyo 182-8585, Japan}
\address{$^{2}$Grid Inc., 107-0061, Tokyo, Japan}
\address{$^{3}$i-PERC, The University of Electro-Communications, 182-8585, Chofu, Tokyo, Japan}
\ead{dsasaki@uec.ac.jp}
\ead{chen.chih.chieh@gridsolar.jp}
\ead{sogabe@uec.ac.jp}

\begin{abstract}
We propose Tree Tensor Network Reservoir Computing (TTN-RC), a quantum-inspired reservoir computing framework for time-series prediction that uses the hierarchical structure of Tree Tensor Networks as a random reservoir. To control the exponential concentration or divergence of TTN outputs, we introduce a hierarchical ensemble method that partitions a fixed-size reservoir into multiple independent sub-reservoirs. In the tested NARMA benchmarks, TTN-RC achieves competitive or improved performance compared with conventional Echo State Networks, especially for tasks requiring higher-order nonlinear processing and longer contextual dependence. We also derive an expected contraction rate based on the reservoir Jacobian and develop a mean-field description of the reservoir-state statistics. These analyses identify an asymptotic stability boundary at $\sigma_{T}=\sqrt{2}$ in the large per-tree-size limit, where several theoretical indicators converge. Our results provide a design principle for tensor-network-based reservoir computing and clarify how hierarchical reservoir topology controls stability and nonlinear information processing.

\end{abstract}

\section{Introduction}

Reservoir computing (RC) is an efficient machine learning paradigm for learning complex temporal dynamics, with prominent implementations like the Echo State Network (ESN) \cite{jaegerEchoStateApproach} and Liquid State Machine (LSM) \cite{maassRealtimeComputingStable2002}. These models have been applied to a broad range of temporal-learning tasks, including time-series forecasting and classification \cite{yanEmergingOpportunitiesChallenges2024, cucchiHandsonReservoirComputing2022}. The core principle of RC involves driving a high-dimensional, nonlinear dynamical system, or ``reservoir,'' with an input signal. The reservoir's internal connections are randomly generated and remain fixed. Consequently, training is simplified to a linear regression that maps reservoir states to the desired outputs. This streamlined process grants RC a significant advantage over traditional Recurrent Neural Networks (RNNs) \cite{Amari1972,elmanFindingStructureTime1990, jordanAttractorDynamicsParallelism1990, Werbos1990, choLearningPhraseRepresentations2014} and Long Short-Term Memory (LSTM) \cite{hochreiterLongShortTermMemory1997} by avoiding issues like vanishing/exploding gradients \cite{Bengio1994} and high computational demands. The concept of RC has inspired a broad range of research, from leveraging physical systems \cite{tanakaRecentAdvancesPhysical2019, nakajimaPhysicalReservoirComputing2020} and quantum dynamics \cite{fujiiHarnessingDisorderedEnsembleQuantum2017, nakajimaBoostingComputationalPower2019, kubotaTemporalInformationProcessing2023, NaturalQuantumReservoir, mujalTimeseriesQuantumReservoir2023, martinez-penaDynamicalPhaseTransitions2021, palaciosRoleCoherenceManybody2024, dudasQuantumReservoirComputing2023, fryOptimizingQuantumNoiseinduced2023, kobayashiFeedbackDrivenQuantumReservoir2024, kobayashiExtendingEchoState2024} to interpretations based on kernel methods \cite{gauthierNextGenerationReservoir2021, gononReservoirKernelsVolterra2022, grigoryevaInfinitedimensionalNextgenerationReservoir2025}.
Among these advancements, Tensor Networks (TNs)—powerful mathematical tools from many-body physics \cite{orusTensorNetworksComplex2019, nishinoCornerTransferMatrix1996, whiteDensityMatrixFormulation1992, eisertColloquiumAreaLaws2010}—have emerged as a promising avenue for machine learning due to their ability to efficiently model complex, high-dimensional data \cite{stoudenmireSupervisedLearningTensor2016, efthymiouTensorNetworkMachineLearning2019, qingCompressingNeuralNetworks2025, rieserTensorNetworksQuantum2023, martinBarrenPlateausQuantum2023, hugginsQuantumMachineLearning2019, liuMachineLearningUnitary2019}.

The structural advantages of TNs were first introduced to RC in a pioneering work on Matrix Product State Reservoir Computing (MPS-RC) \cite{Sato2025}. MPS, a one-dimensional TN, demonstrated that the expressive power of TNs could be harnessed for learning classical nonlinear dynamics. Building on this approach, we propose Tree Tensor Network Reservoir Computing (TTN-RC), which extends the reservoir topology to a Tree Tensor Network (TTN) \cite{Shi2006TTN}. The hierarchical branching structure of a TTN provides a different reservoir topology for representing multi-scale correlation structures. While trainable TTNs have been explored for classification, generation, and model selection \cite{Chen2024TreeClassification, Cheng2019TreeGen, Michel2022TreeModelSelect}, their use as randomized reservoirs has not been systematically explored to our knowledge. Recently, a different approach using TNs for Volterra kernel expansions was proposed \cite{pena2025tntime}, where TNs are employed for compressing time correlations instead of randomizing state-space correlations.

In this work, we propose a \emph{hierarchical ensemble} approach that keeps the total reservoir size $N_x$ fixed while partitioning it into $M$ independent TTNs, each used as a sub-reservoir. This ensemble mitigates the exponential concentration/divergence behavior observed in TTN outputs and provides a practical way to tune the effective tree depth while preserving the total reservoir dimension.
On the theoretical side, we introduce an expected contraction rate $C$ based on the reservoir Jacobian and derive a closed-form expression for it. The condition $C<1$ gives a sufficient contraction criterion in expectation, which serves as an analytical indicator for the ESP-satisfying regime \cite{jaegerEchoStateApproach, buehnerTighterBoundEcho2006, yildizRevisitingEchoState2012, gallicchioChasingEchoState2019}. We further adapt mean-field theory, originally developed in the context of ESNs \cite{massarMeanfieldTheoryEcho2013}, to TTN-RC and examine how the reservoir-state mean square changes with the randomness and scale of the TTN reservoir.
This viewpoint is related to the widely noted phenomenon in reservoir computing that information processing capacity and performance are often maximized near the ``edge of chaos'' \cite{dambreInformationProcessingCapacity2012, kubotaUnifyingFrameworkInformation2021, toyoizumiEdgeChaosAmplification2011}, while our analysis emphasizes an expected stability boundary and mean-field reservoir statistics rather than a deterministic guarantee for each finite random realization.
In addition, we identify an asymptotic phase-boundary behavior with respect to the tensor-element standard deviation $\sigma_T$. In the thermodynamic limit where the per-TTN sub-reservoir size $\tilde{N}_x \to \infty$, the concentration/divergence transition, the value of $\sigma_T$ at which $C=1$, and the mean-field transition point all approach the same critical value $\sigma_T=\sqrt{2}$. This phase-transition-like behavior provides a statistical-physics interpretation of the TTN reservoir topology and a design criterion for selecting reservoir hyperparameters.

Moreover, a scaling analysis near this critical point shows that our model shares some similarities with the previously studied MPS-RC while exhibiting a different dependence on the effective tree size. This behavior originates from the randomized nature of the reservoir, which is related to ideas used in spin-glass-type models \cite{EA1975,SK1975,mezard1986}. A different type of RC phase transition with respect to the randomness of the inputs was discovered in \cite{matsumura2025phase}. Entanglement phase transitions of random TTNs with respect to varying bond dimension are studied in \cite{lopez2020TransitionTTN}.

Beyond its theoretical appeal, we perform benchmark comparisons between ESN and TTN-RC. Configuring TTN-RC with an appropriate ensemble size $M$ to control the exponential concentration/divergence phenomenon, we find that TTN-RC is advantageous in the tested higher-order NARMA tasks, where nonlinear processing and contextual dependence are important.

Our main contributions are as follows:
\begin{itemize}
  \item \textbf{Hierarchical-ensemble TTN reservoir architecture:} partitioning a fixed-size reservoir into multiple independent TTNs to tune the effective tree depth and control concentration/divergence.
  \item \textbf{Empirical performance validation:} demonstrating competitive or improved performance on the tested NARMA tasks, especially in regimes requiring higher-order nonlinear processing.
  \item \textbf{Expected stability characterization:} deriving a closed-form expected contraction rate $C$ and using $C<1$ as an analytical indicator for the ESP-satisfying regime.
  \item \textbf{Asymptotic phase-boundary behavior:} showing that, in the limit $\tilde{N}_x\!\to\!\infty$, three theoretical indicators (concentration/divergence, $C{=}1$, and mean-field reservoir statistics) approach the same critical value $\sigma_T=\sqrt{2}$.

\end{itemize}

The remainder of this paper is organized as follows:
Section~\ref{sec:method} introduces the methods.
Section~\ref{sec:model} formalizes the standard RC setup, reviews ESN, and introduces TTN-RC together with the hierarchical ensemble that partitions a fixed $N_x$ into $M$ independent trees.
Section~\ref{sec:exp} details evaluation metrics ($R^2$, NMSE, $I_{\text{ESP}}$), tasks (short-term memory (STM) and nonlinear autoregressive moving average (NARMA)), and experimental protocols.
Section~\ref{sec:theory} develops our analysis: a Jacobian-based expected contraction rate $C$ that gives an expected stability indicator ($C<1$), together with a mean-field theory for reservoir-state statistics.
Section~\ref{sec:results} presents performance data on STM task and NARMA tasks, and the numerically evaluated ESP index. It also presents head-to-head comparisons with ESN on the NARMA tasks, and numerical evaluations of theoretical measures.
Section~\ref{sec:discussion} analyzes the experimental results with respect to the effects of ensembling and nonlinearity, compares theory to experiments, and offers a discussion of the phase transition and critical behaviors.
Finally, Section~\ref{sec:conclusion} concludes and outlines future directions. The Appendices provide derivations, closed-form formulae, and additional experimental and theoretical details.

\section{Methods}

\label{sec:method}

\subsection{Tree Tensor Network Reservoir Computing}
\label{sec:model}

Reservoir computing is a method for time-series forecasting, classification, and reconstruction using a dynamical system called a ``reservoir.'' The reservoir nonlinearly transforms the incoming data, repeatedly projecting them into a high-dimensional space and maintaining the transformed values as its internal state. The function used to read out the reservoir output, that is, to produce the final output of the model, depends on a set of trainable parameters. Because only the readout-layer parameters are trainable, the training cost is much lower than for RNNs or LSTMs. In the standard RC framework, at each discrete time step $t$, the state-update function $\bm{r}$ and the readout function $\bm{h_\theta}$ with trainable parameters $\bm{\theta}$ are applied as follows:
\begin{align}
    \bm{x}(t) &= \bm{r}(\bm{x}(t-1), \bm{u}(t)) \\
    \bm{y}(t) &= \bm{h_\theta}(\bm{x}(t)) .
\end{align}
Here, $\bm{x}(t)\in \mathbb{R}^{N_x}$ denotes the reservoir state at time $t$, $\bm{u}(t) \in \mathbb{R}^{N_u}$ the input data at time $t$, and $\bm{y}(t) \in \mathbb{R}^{N_y}$ the model's output at time $t$. The state update function $\bm{r}$, called the reservoir, can be instantiated by a wide variety of systems ranging from purely mathematical models to real physical devices. For example, in an Echo State Network (ESN), one uses a sparse random weight matrix together with an activation function. In a standard ESN formulation, the state update and readout are given by:
\begin{align}
    \bm{x}(t) &= \bm{f}(W_{\text{res}} \bm{x}(t-1) + W_{\text{in}} \bm{u}(t)) \\
    \bm{y}(t) &= W_{\text{out}} \bm{x}(t) \ .
\end{align}
In linear regression, $W_\text{out}\in \mathbb{R}^{N_y \times N_x}$ can be quickly estimated using the Moore–Penrose pseudo-inverse of the data matrix \cite{Barata_2011}. In the case of ridge regression, $W_\text{out}$ can be estimated by calculating the inverse matrix given the regularization coefficients. $f$ is a fixed activation function and $\bm{f}$ means its component-wise version. $W_\text{res} \in \mathbb{R}^{N_x \times N_x}$ and $W_\text{in}\in \mathbb{R}^{N_x \times N_u}$ are matrices which are not trainable.

\begin{figure}[t]
\centering
\includegraphics[scale=1.15]{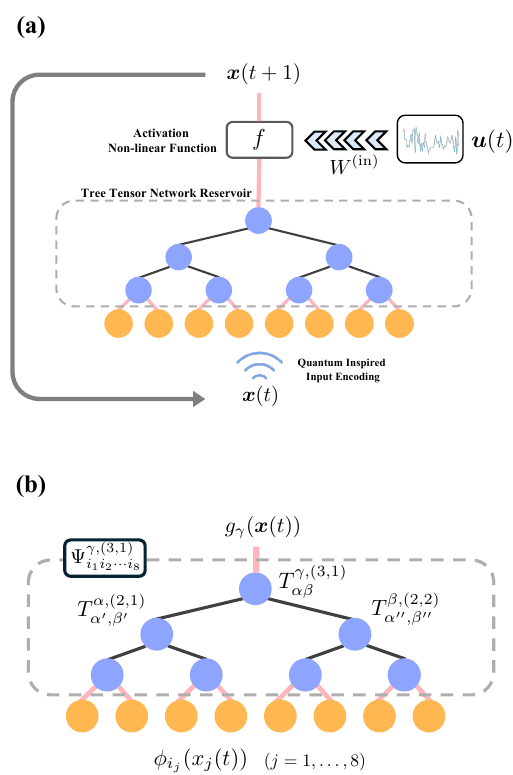}
\caption{Conceptual diagram of TTN-RC ($L=3$):
(a) shows the recursive structure in which the reservoir state is obtained by applying an activation function to the sum of (i) the contraction result of the random TTN function and (ii) the weighted external input, and then feeding that state back as input.
(b) shows only the random TTN function component. Here, the encoded reservoir state at the current time is propagated by successively contracting the random TTN from the bottom up.}
\label{fig:rtt_rc}
\end{figure}

Our proposed model, Tree Tensor Network Reservoir Computing (TTN-RC), is categorized as a quantum-inspired classical reservoir computing model.
In ESN, the reservoir is realized by a sparse random weight matrix; in TTN-RC, a random tree tensor network is used instead. Aside from this difference in how the reservoir state is generated, the linear regression scheme for reading out the output from the reservoir state is identical to that of the ESN. We consider a TTN-RC model with a total reservoir size of $N_x$. This model utilizes $M$ independent TTNs, and their outputs are concatenated. Accordingly, the reservoir size of each TTN is $\tilde{N}_x = N_x / M$. We assume that $M$ divides $N_x$ and that the single-tree size is a power of two. For $M=1$, the tree depth is $L=\log_2 N_x$; for the ensemble case, each tree has depth $\tilde{L}=\log_2 \tilde{N}_x$.

We begin with the case where $M=1$.
In an $L$-layer TTN-RC, the $\gamma$-th element of the reservoir state ($\gamma=1,\ldots,N_x$) is updated as follows:

\begin{align}
    x_\gamma (t+1) &= f(a_\gamma (t)) \label{eq:act_pot0} \\
    a_\gamma (t) &= g_\gamma^{(L)} (\bm{x}(t)) + \sum_{k=1}^{N_u} W_{\gamma k}^\text{(in)} u_k (t)     \label{eq:act_pot1} \\
    g_\gamma^{(L)} (\bm{x}(t)) &= \sum_{i_1, i_2, \dots, i_{N_x}} \Psi_{i_1 i_2 \dots i_{N_x}}^{\gamma, (L,1)} \prod_{j=1}^{N_x} \phi_{i_j} (x_j (t)) .
    \label{eq:ttn_out}
\end{align}
In equation~(\ref{eq:act_pot1}), the first term represents the result of feeding the reservoir state at time $t$ through the random tree tensor network, while the second term corresponds to processing the input data at time $t$ with a random weight matrix. Figure~\ref{fig:rtt_rc}(a) shows a conceptual diagram of the TTN-RC reservoir-update process based on equations~(\ref{eq:act_pot0})--(\ref{eq:ttn_out}). We refer to $a_\gamma(t)$ as the activation potential. $f(x)=\frac{1}{1+e^{-x}}$ is the sigmoid function. Here, for $i_j\in\{0,1\}$, $\phi_{i_j}\bigl(x_j(t)\bigr)$ denotes the variant of the function $\phi_{i_j}$ applied to $x_j(t)$.
In this study, as a specific form of $\phi$,
\begin{equation}
\begin{split}
    \phi_0 (x) &= \cos\left(\dfrac{\pi}{2} x \right) \\
    \phi_1 (x) &= \sin\left(\dfrac{\pi}{2} x \right)
\label{eq:sin-cos}
\end{split}
\end{equation}
is considered.
Here, $x \in [0,1]$, which is automatically satisfied because the sigmoid function is used as the activation function.
In quantum circuits, this encoding is realized using parameterized rotation gates.

Figure~\ref{fig:rtt_rc}(b) shows only the tree tensor reservoir in detail. The tensor network---constructed by contracting three-leg tensors---produces the output via contraction with the input nodes. The tree-tensor operator $\Psi$ is constructed by contracting a tree-structured tensor network. Let $n_l=2^l$ denote the number of input nodes contained in a layer-$l$ subtree. The contraction at layer $l$ can be expressed in terms of the contraction at layer $l-1$ as

\begin{equation}
\Psi_{i_1 i_2 \dots i_{n_l}}^{\gamma,(l,u)}
\;=\;
\sum_{\alpha,\beta}
T_{\alpha\beta}^{\gamma,(l,u)}\,
\Psi_{i_1 i_2 \dots i_{\frac{n_l}{2}}}^{\alpha,(l-1,2u)}\,
\Psi_{i_{\frac{n_l}{2}+1} \dots i_{n_l}}^{\beta,(l-1,2u+1)}.
    \label{eq:ttn_op}
\end{equation}
Each element $T_{\alpha\beta}^{\gamma,(l,u)}$ of the tensor at the $u$-th unit in the $l$-th layer is then sampled once and fixed, according to a zero-mean Gaussian distribution:

\begin{equation}
T_{\alpha\beta}^{\gamma,(l,u)} \;\sim\; \mathcal{N}\ \!\Bigl(0,\;\frac{\sigma_T^2}{d_\alpha d_\beta}\Bigr),
\end{equation}
where $\sigma_T$ is the scale parameter controlling the standard deviation, and $d_\alpha$ and $d_\beta$ denote the dimensions of the bond-index spaces associated with $\alpha$ and $\beta$, respectively. In the simulations below, all internal bond-index spaces have the common bond dimension $\chi$, so $d_\alpha=d_\beta=\chi$ for internal bond legs; at the first layer, the physical input indices $i_j\in\{0,1\}$ have dimension 2.
This type of normalization, where the variance is divided by the bond dimension, is inspired by research on random weight initialization in neural networks \cite{jacotNeuralTangentKernel2020, pooleExponentialExpressivityDeep2016, schoenholzDeepInformationPropagation2017, karakidaUniversalStatisticsFisher2019}. This approach helps mitigate the model's dependency on the bond dimension in theoretical analyses.

Next, we describe the ensemble case for $M\ge 2$. Each tree has size $\tilde{N}_x$ and depth $\tilde{L}$, and the local reservoir index is $\gamma=1,\ldots,\tilde{N}_x$:

\begingroup

\begin{align}
    x_\gamma^{(m)} (t+1) &= f(a_\gamma^{(m)} (t)) \\
    a_\gamma^{(m)} (t) &= g_\gamma^{(\tilde{L},m)} (\bm{x}^{(m)}(t)) + \sum_{k=1}^{N_u} W_{\gamma k}^{\text{(in)},m} u_k (t)     \label{eq:act_pot2} \\
    g_\gamma^{(\tilde{L},m)} (\bm{x}^{(m)}(t)) &= \sum_{i_1, i_2, \dots, i_{\tilde{N}_x}} \Psi_{i_1 i_2 \dots i_{\tilde{N}_x}}^{\gamma, (\tilde{L},1,m)} \prod_{j=1}^{\tilde{N}_x} \phi_{i_j} (x_j^{(m)} (t)) .
    \label{eq:ttn_out2}
\end{align}
\endgroup
\begin{equation}
\Psi_{i_1 i_2 \dots i_{n_l}}^{\gamma,(l,u,m)}
\;=\;
\sum_{\alpha,\beta}
T_{\alpha\beta}^{\gamma,(l,u,m)}\,
\Psi_{i_1 i_2 \dots i_{\frac{n_l}{2}}}^{\alpha,(l-1,2u,m)}\,
\Psi_{i_{\frac{n_l}{2}+1} \dots i_{n_l}}^{\beta,(l-1,2u+1,m)}.
    \label{eq:ttn_op2}
\end{equation}
\begin{equation}
T_{\alpha\beta}^{\gamma,(l,u,m)} \;\sim\; \mathcal{N}\ \!\Bigl(0,\;\frac{\sigma_T^2}{d_\alpha d_\beta}\Bigr),
\end{equation}
\begin{equation}
 \bm{x}(t+1) = [\bm{x}^{(1)}(t+1);\bm{x}^{(2)}(t+1); \dots ;\bm{x}^{(M)}(t+1)],
\end{equation}
where the superscript $m = 1,2,\dots,M$ is an index used to distinguish independent trees.

The tensor elements and input weights of different trees are drawn independently from the same distributions. In this respect, each tree independently enables slightly different pathways of information propagation.
Extensions of this approach could include ensembling trees of varying depths or, more generally, ensembling trees drawn from entirely separate random processes.
Such an ensemble technique is analogous to spatial multiplexing in quantum reservoir computing \cite{nakajimaBoostingComputationalPower2019}.

\begin{figure}[t]
\centering
\includegraphics[scale=0.8]{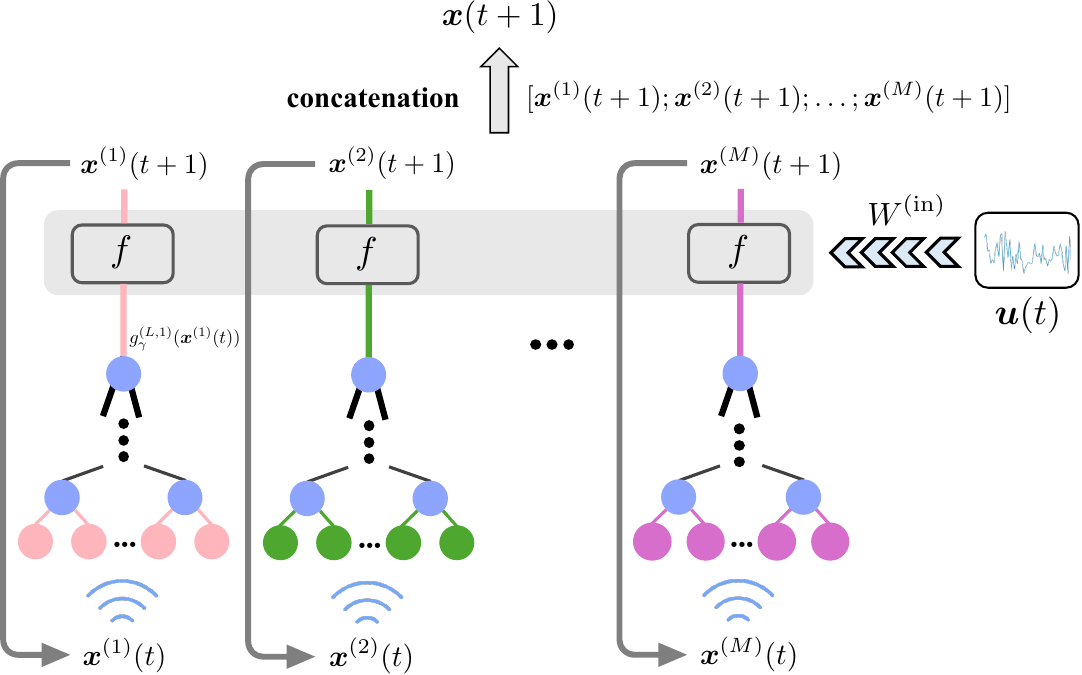}
\caption{ The recursive structure of the ensemble TTN-RC. Each tree is contracted and recursed independently, but their outputs are concatenated to form the overall reservoir state vector.}
\label{fig:ens}
\end{figure}

\subsection{Experimental Methods}
\label{sec:exp}

In supervised learning with reservoir computing, we consider a set of input and target output pairs $\{u(t), d(t)\}_t$. At each time step, the input $u(t)$ is fed into the reservoir to generate the reservoir state $\bm{x}(t)$. Subsequently, the output weights $W_\text{out}$ are trained such that the output $y(t)$ approximates the desired output $d(t)$.

To evaluate the performance of the time-series regression model, we utilize the coefficient of determination ($R^2$ score) and the normalized mean squared error (NMSE). These metrics quantify the accuracy of the predicted series $y(t)$ with respect to the true series $d(t)$ over a discrete time period with $T$ data points.

The $R^2$ score measures the proportion of the variance in the true data that is explained by the model's predictions. It is defined as:
\begin{equation}
    R^2 = 1 - \frac{\sum_{t=1}^{T} (d(t) - y(t))^2}{\sum_{t=1}^{T} (d(t) - \bar{d})^2}
    \label{eq:r2}
\end{equation}
where $\bar{d}$ is the mean of the true series, calculated as $\bar{d} = \frac{1}{T} \sum_{t=1}^{T} d(t)$. An $R^2$ score closer to 1.0 indicates a better fit, signifying that the model's predictions closely match the actual data.

The NMSE is the mean squared error normalized by the variance of the true series. This normalization makes the metric independent of the data's scale. It is calculated as:
\begin{equation}
    \text{NMSE} = \frac{\sum_{t=1}^{T} (d(t) - y(t))^2}{\sum_{t=1}^{T} (d(t) - \bar{d})^2}
    \label{eq:nmse}
\end{equation}
A lower NMSE value indicates a more accurate prediction, with 0 being a perfect score. From equations \eqref{eq:r2} and \eqref{eq:nmse}, it is clear that the two metrics are related by $R^2 = 1 - \text{NMSE}$.

To numerically verify the ESP, we introduce the metric $I_\text{ESP}(t)$:
\begin{equation}
I_\text{ESP}(t) = \frac{\|\bm{x}(t) - \bm{x'}(t)\|}{\|\bm{x}(0) - \bm{x'}(0)\|} .
\end{equation}
Here, $\bm{x}(t)$ and $\bm{x'}(t)$ are the reservoir states at time $t$, resulting from state updates repeated up to time $t$ in the same reservoir, starting from different initial states $\bm{x}(0)$ and $\bm{x'}(0)$, respectively. In other words, $I_\text{ESP}(t)$ is a metric that indicates how much the difference between the reservoir states at time $t$ has decreased (or increased) relative to the initial difference. We consider the ESP to be numerically satisfied if this metric decays from 1 at time 0 and converges to a sufficiently small value at time $t$.
The initial states $\bm{x}(0)$ and $\bm{x}'(0)$ were sampled from a uniform distribution on $[0,1]$. In our experiments, we fixed this pair of initial states and calculated the average value of $I_\text{ESP}(100)$ over 10 different reservoir realizations.

To evaluate the reservoir's ability to retain information from past inputs, we utilize the short-term memory (STM) task.
This task quantifies how well the reservoir can reconstruct a past input signal $u(t-\tau_d)$ from the current reservoir state $\bm{x}(t)$ for various time delays $\tau_d$.
The input signal $u(t)$ is independently sampled at each discrete time step $t$ from a uniform distribution over the interval $[-1, 1]$.
For each specific delay $\tau_d$, the reservoir's prediction $y(t, \tau_d)$ is generated to approximate the target signal $u(t-\tau_d)$.
The performance for each delay $\tau_d$ is measured by $R^2$ score between the target and the predicted signal through linear regression. We denote this score as the STM capacity $C_\text{STM}(\tau_d)$.
A value of $C_\text{STM}(\tau_d)$ close to 1 indicates a near-perfect reconstruction of the input with delay $\tau_d$.
To obtain a single metric for the overall memory performance of the reservoir, we define the total STM capacity, $C_\text{STM}^\text{tot}$, as the sum of the capacities for all delays up to a predefined maximum delay, $\tau_{\text{max}}$:
\begin{equation}
  C_\text{STM}^\text{tot} = \sum_{\tau_d=1}^{\tau_{\text{max}}} C_\text{STM}(\tau_d)
\end{equation}
This cumulative measure reflects the total amount of information about the past input history that is linearly accessible from the current state of the reservoir.

The Nonlinear Auto-Regressive Moving Average (NARMA) task is a standard benchmark for evaluating time-series prediction models, particularly their ability to learn long-term, nonlinear dynamics. In this study, we utilize the NARMA-n sequence, a variant that allows for the systematic assessment of a model's memory capacity and nonlinear processing capabilities by varying the order $n$. The sequence is formally defined by the following recurrence relation:
\begin{equation}
d(t) = \alpha d(t-1) + \beta d(t-1) \sum_{i=1}^n d(t-i) + \gamma u(t-n) u(t-1) + \delta .
\end{equation}
where $d(t)$ is the output at time step $t$, and $u(t)$ is an input drawn from a uniform distribution on $[0, 0.5]$. Following established conventions, the parameters are set to $(\alpha, \beta, \gamma, \delta) = (0.3, 0.05, 1.5, 0.1)$.
The NARMA-$n$ task presents two primary challenges for any learning model. First, since the calculation of $d(t)$ depends on the history of both outputs $\{d(t-1), \dots, d(t-n)\}$ and inputs $\{u(t-1), \dots, u(t-n)\}$, a successful model must possess the capacity to maintain long-term dependencies. Second, the presence of product terms, such as $d(t-1) \sum_{i=1}^n d(t-i)$ and $u(t-n)u(t-1)$, introduces significant nonlinearity. Consequently, the model must have a strong capability for nonlinear transformation to accurately capture the underlying data-generating process.
The performance on the NARMA tasks is evaluated in terms of the $R^2$ score and the NMSE.

\subsection{Theoretical analyses}
\label{sec:theory}

To understand the learning behavior of TTN-RC, we conduct theoretical and numerical analyses based on the statistical properties of the TTNs. We independently develop an expected Lipschitz analysis and a mean-field theory (MFT) \cite{massarMeanfieldTheoryEcho2013}.

The expected Lipschitz constant for the random reservoir \cite{Sato2025}, defined by $\langle \| \bm{x}_1(t+1)-\bm{x}_2(t+1) \| \rangle_T  \le C \| \bm{x}_1(t)-\bm{x}_2(t) \|$, is calculated through a Jacobian matrix analysis, $C=\frac{\tilde{N}_x}{4}\sqrt{V_J^{(l)}}$ where $V_J^{(l)}=\left\langle \left(\frac{\partial g_{\gamma}^{(l)}(\bm{x})}{\partial x_{m}}\right)^2 \right\rangle_T$.
The MFT here relies on the large bond dimension limit and a stationary-state approximation, while the Lipschitz analysis does not require these assumptions. The input-driven term is ignored in this mean-field calculation, so the resulting quantities should be regarded as analytical reservoir statistics rather than a full description of the input-driven dynamics. These approximations allow us to calculate the statistics of the current reservoir state, since the pre-activation state can be approximated by a normal distribution.
In the case of ESN, the variance of the current reservoir state explicitly depends on the reservoir state of the previous time step, and hence a self-consistent calculation is required to obtain the stationary distribution properties. In the case of MPS-RC, the mean-field stationary statistics depend on both the internal state $\bm{x}$ and the bond dimension $\chi$. Fortunately, in TTN-RC, $\langle (g_\gamma^{(l)} (\bm{x}))^2 \rangle_T$ is independent of $\bm{x}$ and independent of $\chi$.
For simplicity, we ignore the input-driven term and calculate the mean-field moments

\begingroup

\begin{align}
q_x := \left\langle x_\gamma(t)^2 \right\rangle
&= \int \mathcal{D}a \ f^2(a) \label{eq:qx} \\
\mu_x := \left\langle x_\gamma(t) \right\rangle &= \int \mathcal{D}a \ f(a) \\
v_x := \operatorname{Var}(x_\gamma(t)) &= q_x-\mu_x^2 ,
\label{eq:MFVar}
\end{align}
\endgroup
 $\forall t, \forall \gamma$. For the sigmoid activation and the zero-mean Gaussian measure below, $\mu_x=1/2$, so $v_x=q_x-1/4$. We use $q_x=\langle x_\gamma^2\rangle$ for the mean-field heatmap in Fig.~\ref{fig:theory}(b), and $v_x$ for the green variance curves in Fig.~\ref{fig:mc_narma_heatmap}(e,f). The Gaussian integral measure is defined by

\begin{equation}
    \mathcal{D}a := \frac{da}{\sqrt{2 \pi V_g^{(\tilde{L})}}} \exp\left[- \frac{a^2}{2 V_g^{(\tilde{L})}}\right],
\end{equation}
where $V_g^{(\tilde{L})}$ is the variance given by
\begin{equation}
    V_g^{(\tilde{L})} = \bigl\langle \left(g_\gamma (\bm{x}(t))\right)^2 \bigr\rangle_T = \frac{(\sigma_T^2)^{\tilde{N}_x-1}}{2^{\tilde{N}_x}} ,
\end{equation}
and $\tilde{L} := \log_2 {\tilde{N}_x}$ represents the depth of each tree. A critical value $\sigma_T=\sqrt{2}$ can be extracted under the large $\tilde{N}_x$ limit. The mean-field calculations are used to interpret stability-related indicators \cite{massarMeanfieldTheoryEcho2013} and to construct a phenomenological learning-theory proxy \cite{gononRiskBoundsReservoir2020}.

Further theoretical details are provided in the appendices. The Jacobian matrix analysis is presented in \ref{subsec:ESP}. The mean-field learning-theory phenomenology is presented in \ref{subsec:GGO}, the scaling analysis near criticality in \ref{subsec:scaling_apndx}, and the ESP-index and contraction analyses in \ref{subsec:ESPindex}. Calculation details are provided in \ref{subsec:calculations}.

\section{Results}
\label{sec:results}

\subsection{Comparison with ESN}
\label{sec:namra_comparison}

\begin{figure}
\centering
\includegraphics[width=1.0\linewidth]{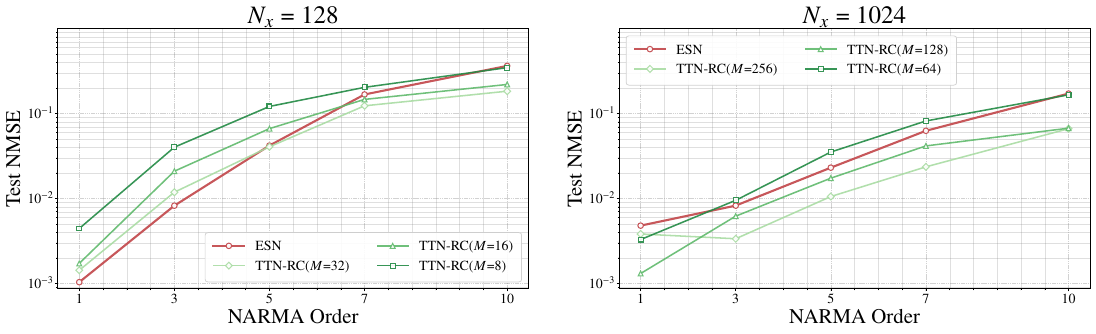}
\caption{Comparison of the performance between TTN-RC and ESN on the NARMA1, 3, 5, 7, and 10 tasks. The horizontal axis represents the order of the NARMA task, $n=1,3,5,7,10$, and the vertical axis represents the NMSE. In the left panel, the total reservoir size is fixed at $N_x=128$, while in the right panel, it is fixed at $N_x=1024$. For TTN-RC, the results are shown for values of $M$ corresponding to $\tilde{N}_x=4, 8, \text{and } 16$.
The presented result is the NMSE on the test data, averaged over 10 different reservoir realizations, using optimized parameters. Outliers, defined as data points falling outside 1.5 times the interquartile range, were removed, after which the average NMSE was calculated.
}
\label{fig:narma_optuna_NMSE}
\end{figure}

We compared the performance of TTN-RC and ESN on the NARMA1, 3, 5, 7, and 10 tasks using the NMSE. The lengths of the training data, validation data (for hyperparameter optimization), and test data were set to 3000, 500, and 500, respectively. A washout period of 500 steps was applied before the training data. The hyperparameters for both ESN and TTN-RC were optimized using Optuna \cite{akiba2019optunanextgenerationhyperparameteroptimization}. The sigmoid function was uniformly used as the activation function $f$ for both models.

The results of this comparison are shown in Fig.~\ref{fig:narma_optuna_NMSE}. In the left panel, the total reservoir size is fixed at $N_x=128$, and in the right panel, it is fixed at $N_x=1024$. For TTN-RC, the plots show the NMSE for models partitioned into $M$ ensembles, corresponding to $\tilde{N}_x=4, 8, \text{and } 16$. In these optimized runs, ESN is comparable to or better than TTN-RC on the lower-order NARMA tasks for $N_x=128$, while the relative performance of TTN-RC improves as the task order increases. For $N_x=1024$, TTN-RC gives lower NMSE than the ESN baseline across the tested task orders. Furthermore, although the task difficulty increases with higher-order NARMA tasks, the performance degradation of TTN-RC is more gradual in these tests, suggesting improved robustness in the tested settings.

\subsection{Memory Capacity, Performance, and ESP}
\label{sec:result_performance}

\begin{figure}
\centering
\includegraphics[width=0.92\linewidth]{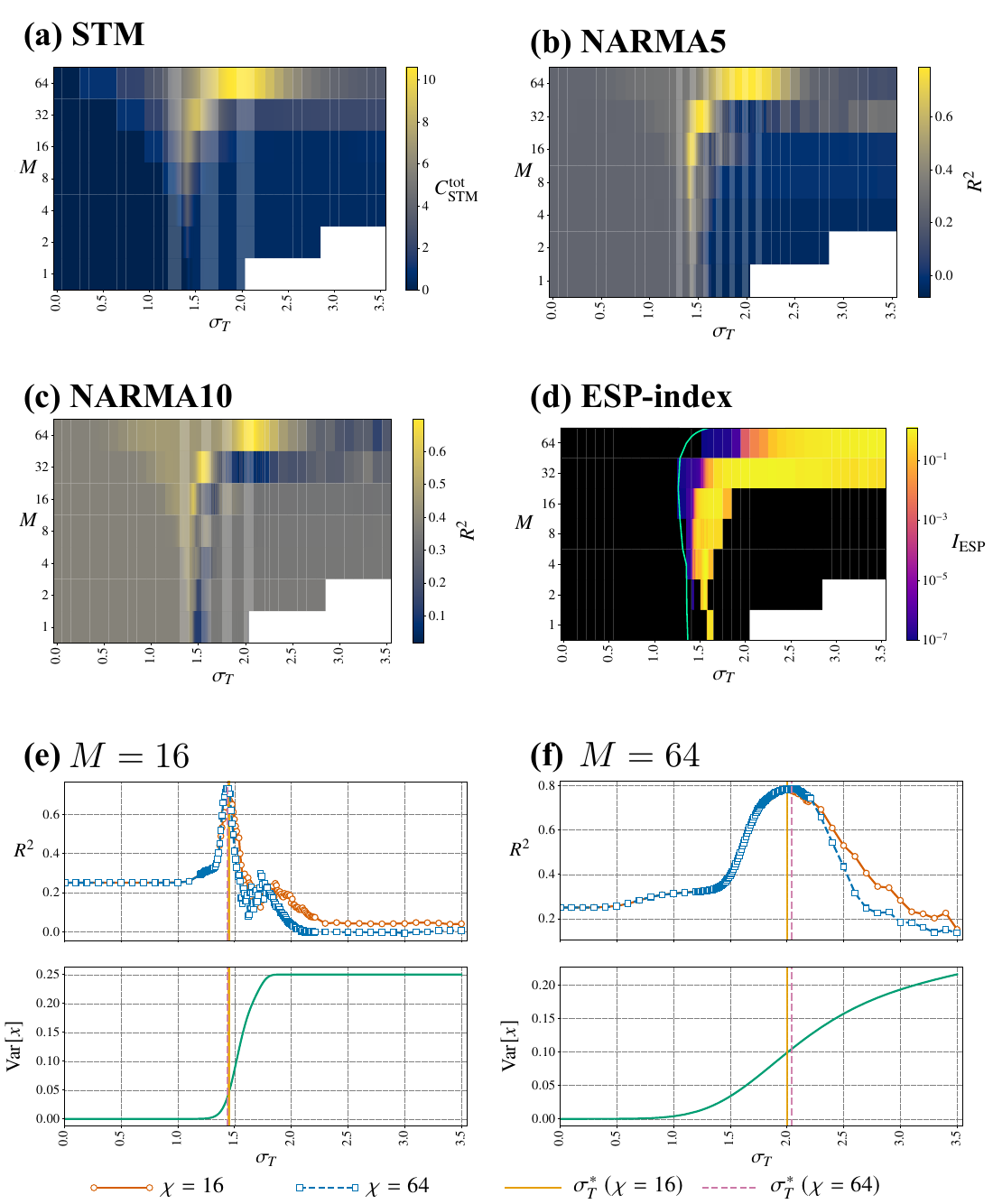}
\caption{ Performance evaluation of the TTN-RC.
        \textbf{(a):} Heatmap of the total memory capacity, $C_{\text{STM}}^{\text{tot}}$, for the STM task.
        \textbf{(b, c):} Heatmaps of the $R^2$ score for the NARMA5 and NARMA10 tasks, respectively.
        \textbf{(d):} Heatmap of the ESP metric, $I_{\text{ESP}}(100)$. The light-green line indicates the contour for $C=1$. In panel (d), values of $10^{-7}$ or less are displayed in the same color.
        All metrics presented in the figures are averaged over 10 different reservoir realizations. The white regions in the heatmaps indicate parameter settings where the TTN computation diverged, making subsequent calculations impossible.
         \textbf{(e, f):} The $\sigma_T$-dependence of the $R^2$ score on the NARMA5 task for a fixed $M$. The $R^2$ scores were calculated for $\chi=16$ and $64$, and the optimal $\sigma_T = \sigma_T^\ast$ that yields the best performance is indicated for each $\chi$ (solid yellow line and dashed pink line, respectively).
         The green curve represents the mean-field reservoir-state variance $v_x=\operatorname{Var}(x_\gamma)=q_x-\mu_x^2$ as a function of $\sigma_T$.
}
\label{fig:mc_narma_heatmap}
\end{figure}

We evaluated the performance of the TTN-RC on several benchmark tasks.
Here, the total reservoir size was fixed at $N_x=256$.
Specifically, we calculated the total memory capacity, denoted as $C_{\text{STM}}^{\text{tot}}$, for the STM task. Additionally, we computed the $R^2$ scores for the NARMA5 and NARMA10 tasks. For all simulations, the input weights $W_{\text{in}}$ were sampled from a uniform distribution on $[-1, 1]$, and the bond dimension was set to $\chi=64$.

For the STM task, the training and test data lengths were set to 4000 and 1000, respectively, with a washout period of 100 steps. The total memory capacity was calculated by summing the memory capacity $C_{\text{STM}}(\tau_d)$ up to a maximum delay of $\tau_{\text{max}}=100$. For computational convenience, negative values of $C_{\text{STM}}(\tau_d)$ were replaced with 0 before summation. For the NARMA tasks, the training and test data lengths were 900 and 100, respectively, with a washout period of 50 steps. A regularization coefficient of $\lambda=1.0$ was used for the ridge regression readout.

The resulting heatmaps for $C_{\text{STM}}^{\text{tot}}$ and the $R^2$ scores for NARMA5 and NARMA10 are shown in Fig.~\ref{fig:mc_narma_heatmap}(a-c). A clear trend is observed where both memory capacity and prediction accuracy improve as the number of ensembles $M$ increases. Furthermore, the area of this high-performance region expands with increasing $M$.

To further investigate the dynamical properties of the reservoir, we use the ESP metric, $I_{\text{ESP}}(100)$, which measures the system's sensitivity to different initial states.
For this calculation, the input $u(t)$ was sampled independently at each time step from a uniform distribution on $[-1, 1]$. The resulting heatmap, shown in Fig.~\ref{fig:mc_narma_heatmap}(d), visualizes this property across the parameter space $(\sigma_T, M)$.

For $M=1,2,4,8,16$, as $\sigma_T$ was increased, re-entrant behavior was observed in which the system transitioned from a regime with a small ESP index to a regime with a large ESP index and then returned to a small-index regime. However, for $M=32,64$, such behavior was not observed within the parameter range of this study. To the left of the $C=1$ line (light-green line), the condition $I_\text{ESP}(100)\leq 10^{-7}$ is met, which is consistent with $C<1$ acting as an expected stability indicator. The Jacobian-based expected-contraction analysis alone cannot explain the reappearance of the small-ESP-index regime, and a mean-field theoretical approach will be discussed.

A comparison between Fig.~\ref{fig:mc_narma_heatmap}(a-c) and Fig.~\ref{fig:mc_narma_heatmap}(d) reveals that for a fixed $M$, the performance is maximized during the transition of $I_\text{ESP}(100)$ from small values (the black region) to large values (the yellow region).

Figs.~\ref{fig:mc_narma_heatmap}(e) and (f) show the evolution of the $R^2$ score for the NARMA5 task as a function of $\sigma_T$ for $M=16,64$, to verify that the region of maximum performance is contained within the slope region of the mean-field reservoir-state variance $v_x$. The reservoir and task settings are identical to those for the data in the $R^2$ heatmap for NARMA5, but we also display the data for $\chi=16$. It is evident that the line for $\sigma_T^\ast$, which maximizes performance, is located within the slope region of $v_x$. Furthermore, although there are minor differences in performance and the value of $\sigma_T^\ast$ depending on $\chi$, for both $\chi=16$ and $64$, the optimal $\sigma_T^\ast$ falls within this slope region.

\subsection{Theoretical Metrics}
\label{subsec:eoc_results}

\begin{figure}
\centering
\includegraphics[width=\linewidth]{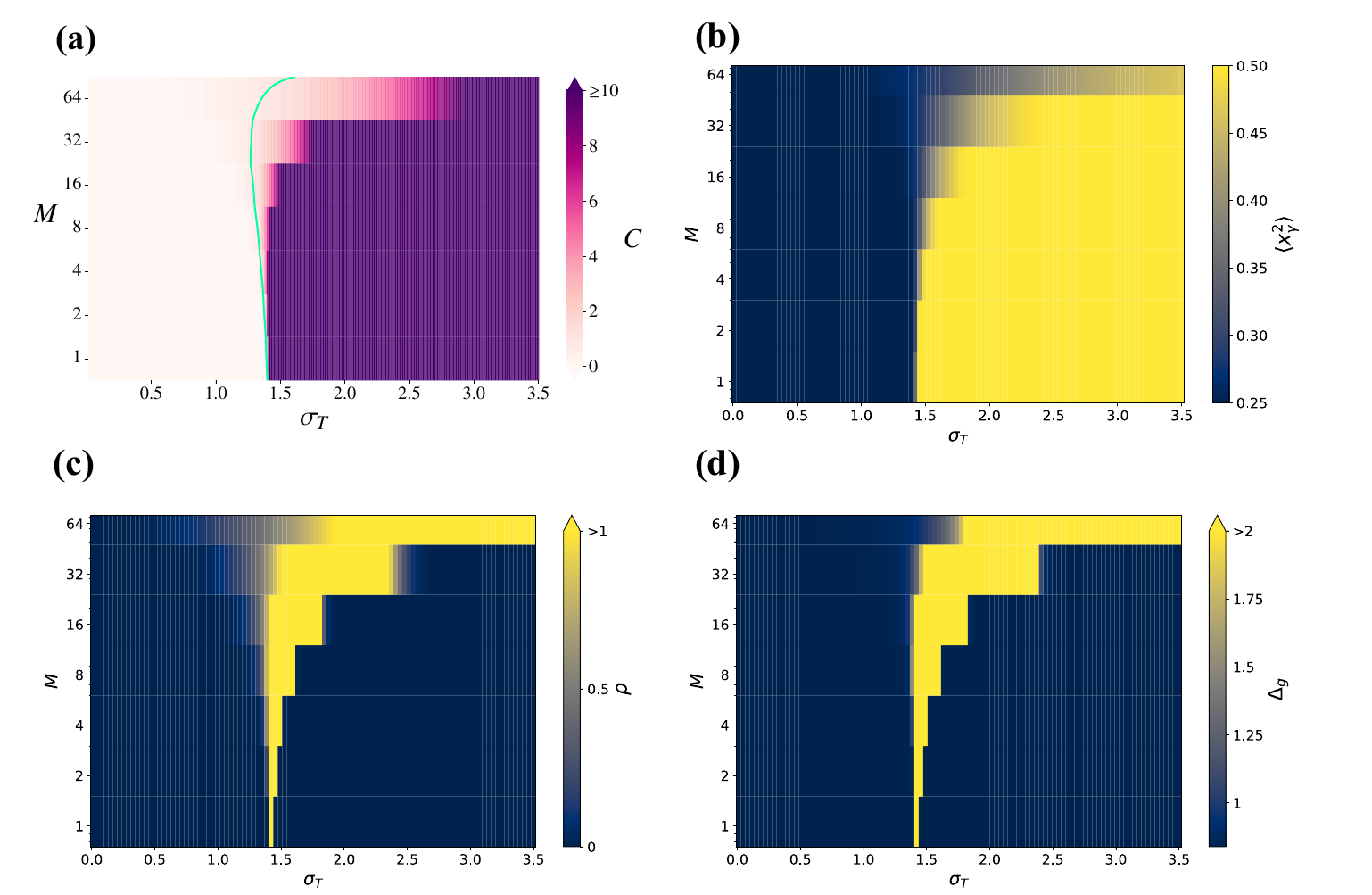}
\caption{Heatmaps of theoretical metrics. The horizontal axis represents $\sigma_T$, and the vertical axis represents the number of trees $M$. $N_x=256$ for all calculations. \textbf{(a):} Heatmap of $C$, obtained by calculating the reservoir's Jacobian. Values of $C \geq 10$ are saturated to the same color. The light-green line indicates the contour for $C=1$.
\textbf{(b):} Heatmap of the reservoir-state mean square $q_x=\langle x_\gamma^2\rangle$, given by equation \eqref{eq:qx}, calculated using mean-field theory. \textbf{(c):} Mean-field contraction constant $\rho$. \textbf{(d):} Phenomenological learning-theory proxy based on a mean-field generalization-gap bound.}
\label{fig:theory}
\end{figure}

Fig.~\ref{fig:theory} shows several different theoretical indicators. We calculate the expected Lipschitz constant of the TTN-RC, which provides a sufficient contraction condition in expectation and serves as an analytical indicator for the ESP-satisfying regime. Details are provided in \ref{subsec:ESP}.
This condition can be expressed using the standard deviation $\sigma_T$ governing the randomness of the tensors and the reservoir dimension $\tilde{N}_x = N_x / M$ of each tree, as follows:
\begin{equation}
C := \frac{\pi N_x}{8\sqrt{2}M}\left(\frac{\sigma_T}{\sqrt{2}}\right)^{\frac{N_x-M}{M}} < 1 .
\end{equation}
The satisfaction of $C<1$ indicates an expected contraction regime. If $C\geq1$, this expected-contraction condition no longer holds, and the reservoir dynamics can become sensitive to initial conditions. The condition $C=1$ is therefore used as an analytical estimate of the stability boundary.
Such $\sigma_T$ is given by
\begin{equation}
\sigma_{T}^{(C=1)} = \sqrt{2} \left(\frac{8\sqrt{2}M}{\pi N_x}\right)^{\frac{M}{N_x -M}}.
\label{eq:jac_esp_C}
\end{equation}
Fig.~\ref{fig:theory}(a) provides a heatmap of the expected Lipschitz constant $C$. For $C<1$, the system is in the expected-contraction regime, while $C>1$ means that this expectation-level contraction condition is no longer available. The light-green line is the $C=1$ contour line. $C$ increases with $\sigma_T$, starting from $C\approx 0$ and reaching large values $C>10$, with values larger than 10 truncated. The growth of $C$ differs among values of $M$. For small $M \rightarrow 1$ (single large tree), $C$ is almost constant for $\sigma_T<\sqrt{2}$, and a sharp transition from small $C$ to large $C$ can be observed near the critical value $\sigma_T=\sqrt{2}$. For large $M$, the growth is smoother, and a crossover regime emerges. The contour line for $C=1$ also deviates from the critical value $\sigma_T=\sqrt{2}$ at large $M$. The crossover regime becomes wider and shifts toward $\sigma_T>\sqrt{2}$ for large $M$. The consistency with the experimental results will be demonstrated later.

Fig.~\ref{fig:theory}(b) provides a heatmap of the MFT reservoir-state mean square $q_x=\langle x^2_\gamma \rangle$. Details are provided in \ref{subsec:calculations}. Although this calculation is independent of that for Fig.~\ref{fig:theory}(a), the heatmaps appear similar. $q_x$ increases with $\sigma_T$ from $q_x \approx 0.25$ to $q_x \approx 0.5$, without any truncation. There is a broad regime for $q_x \approx 0.25$ and another for $q_x \approx 0.5$, revealing two operating regimes. There is a sharp transition-like change at $\sigma_T=\sqrt{2}$ for small $M$, while for large $M$ the crossover is smooth. The crossover becomes wider and shifts toward $\sigma_T>\sqrt{2}$ for large $M$.

Based on the calculations in Figs.~\ref{fig:theory}(a) and (b), the heatmap in Fig.~\ref{fig:theory}(c) is derived for the mean-field contraction constant $\rho$. Details are provided in \ref{subsec:ESPindex}. The yellow regime for $\rho\rightarrow 1$ indicates a mean-field sensitive regime, while the black regime for $\rho\rightarrow 0$ indicates a mean-field stable regime. For both large and small $\sigma_T$, the system is stable in this mean-field sense, while a sensitive regime appears in between. The shape and location of the sensitive regime resemble the crossover regime in Figs.~\ref{fig:theory}(a) and (b). For small $M$, this regime converges to a small area near the critical value $\sigma_T=\sqrt{2}$, while for large $M$ the area becomes wider and shifts toward larger $\sigma_T$.

Fig.~\ref{fig:theory}(d) provides an evaluation of a phenomenological learning-theory proxy. Details are provided in \ref{subsec:GGO}. The heatmap is based on a mean-field evaluation of the generalization-gap upper bound $\Delta_g$. In the present manuscript, $\Delta_g$ is used as an interpretive indicator rather than as a quantitative predictor of the observed test error. Similar to the sensitive regime in Fig.~\ref{fig:theory}(c), $\Delta_g$ is low in the two mean-field regimes and high in the crossover regime of Figs.~\ref{fig:theory}(a) and (b). The high-$\Delta_g$ regime is sharp and close to the critical value $\sigma_T=\sqrt{2}$ for small $M$, whereas it is broader and smoother for large $M$. The regime shifts toward larger $\sigma_T$ for large $M$.

\section{Discussion}
\label{sec:discussion}

\subsection{Better performance over ESNs}

In the tested NARMA settings, TTN-RC showed advantages over the ESN baseline in several higher-order tasks. This result suggests a potential benefit of adopting a hierarchical TN reservoir topology for predicting time series with long-range correlations. Although this study focused on a comparison between TTN-RC and the ESN baseline, a detailed comparative analysis of the performance and computational costs among different tensor network structures on various tasks remains an important topic for future work.

\subsection{Tree Depth, Nonlinearity, Stability, and Optimal Hyperparameters}

For the stability and ESP diagnostics, the ESP-index data in Fig.~\ref{fig:mc_narma_heatmap}(d) are consistent with both the expected-contraction indicator from the Jacobian matrix analysis (Fig.~\ref{fig:theory}(a)) and the MFT contraction constant (Fig.~\ref{fig:theory}(c)). The large ESP-index regime is located on the right side of the $C=1$ contour line, where the left edge closely follows the shape of the contour line. The large ESP-index regime also resembles the high-sensitivity regime predicted by the mean-field contraction constant. The MFT predicts the location of this regime at the crossover of two mean-field operating regimes, indicated by the mean-square order parameter (Fig.~\ref{fig:theory}(b)).

As shown in Fig.~\ref{fig:mc_narma_heatmap} (e) and (f), the range of $\sigma_T$ that maximizes the performance for a fixed $M$ can be identified by the slope region of the variance $v_x$. However, the change in performance due to the variation in $M$ is not fully explained by our theory.

Even if the behaviors of the stability indicators shown in Fig.~\ref{fig:theory}(c) and Fig.~\ref{fig:mc_narma_heatmap}(d) are similar, the memory capacity and task performance presented in Fig.~\ref{fig:mc_narma_heatmap}(a-c) differ.
For instance, in the upper-right corner ($M=64,\sigma_T\in[2.5,3.5]$), both the theoretical $\rho$ and experimental $I_{\text{ESP}}$ heatmaps show high values, but the STM and NARMA5 task performances are low.
This implies that stability alone is not sufficient for specifying the parameter region that maximizes the time-series modeling performance of TTN-RC in the tested settings. The sloping area of the theoretical heatmap $\Delta_g$ in Fig.~\ref{fig:theory}(d) may provide an additional phenomenological indicator for the performance peaks of STM and NARMA in Fig.~\ref{fig:mc_narma_heatmap}(a-c), but the calculation relies on simplifying assumptions and should not be regarded as a quantitative prediction of the test error. The decomposed data for the training error and generalization gap can be found in \ref{subsec:generalize}. It would be desirable to also include a theory for the approximation error \cite{YASUMOTO2025}, but this is beyond the scope of the current work. Intuitively, there are two considerations for choosing the size (or equivalently, the number) of the trees. First, when the trees are deep, the crossover regime becomes so sharp that it is difficult to locate the optimal parameter (see Section~\ref{subsec:criticality} and \ref{subsec:scaling_apndx} for the scaling analysis). Second, it is desirable to adjust the nonlinearity and memory of the reservoir according to the task \cite{matsumura2025phase}. In our study, the depth of the tree is likely the primary factor adjusting the nonlinearity. As the tree becomes deeper, the number of products during the contraction of the tensor network increases, which is thought to result in a reservoir that generates a very high-order polynomial. We therefore empirically speculate that TTN-RC with depths corresponding to approximately $\tilde{N}_x=4, 8$ (corresponding to $M=64,32$) was effective for the tested NARMA tasks.

\subsection{Exponential concentration/divergence, phase transitions, and quantum RC}

\label{subsec:criticality}

The mean and variance of the output from the TTN with respect to the randomness of the tensor $T$ can be analytically calculated:
\begin{align}
    \bigl\langle g_\gamma (\bm{x}(t)) \bigr\rangle_T &= 0 \\
    \bigl\langle \left(g_\gamma (\bm{x}(t))\right)^2 \bigr\rangle_T &= \frac{(\sigma_T^2)^{\tilde{N}_x-1}}{2^{\tilde{N}_x}} .
\end{align}
By analogy with statistical physics, we consider the thermodynamic limit $\tilde{N}_x \rightarrow \infty$.
The variance $\bigl\langle \left(g_\gamma (\bm{x}(t))\right)^2 \bigr\rangle_T$ diverges for $\sigma_T > \sqrt{2}$ and converges to zero for $\sigma_T < \sqrt{2}$, with $\sigma_T = \sqrt{2}$ as the critical point.
This likely corresponds to cases where the reservoir state $\bm{x}$ either exponentially concentrates around its mean value or becomes too random to retain information from the input.
This implication highlights the importance of keeping $\tilde{N}_x = N_x/M$ at an appropriate size.
The ensemble number $M$ plays a crucial role in preventing $\tilde{N}_x$ from becoming too large while maintaining the total reservoir size $N_x$.
Ensembling is therefore considered effective in this respect.
The exponential concentration problem in quantum reservoir computing has been discussed in recent studies \cite{sannia2025exponentialconcentrationsymmetriesquantum, xiong2025rolescramblingnoisetemporal}.
Furthermore, the value of $\sigma_T$ that satisfies $C=1$ in Eq.~\eqref{eq:jac_esp_C} also converges to $\sqrt{2}$.

In Fig.~\ref{fig:theory}(b), it can be observed that $q_x$ undergoes a sharp transition-like change around $\sigma_T = \sqrt{2}$ for small values of $M$ (i.e., for large $\tilde{N}_x$).
Thus, multiple indicators suggest that in the thermodynamic limit $\tilde{N}_x \rightarrow \infty$, $\sigma_T=\sqrt{2}$ describes an asymptotic transition between different reservoir operating regimes.

This supports our observation that the maximum performance for a finite $\tilde{N}_x$ is achieved in the vicinity of $\sigma_T=\sqrt{2}$.
In MPS-RC, the only phase transition point in the thermodynamic limit was at $\sigma_T=0$.

By analogy with statistical physics, this behavior can be interpreted as a finite-temperature-like transition in the TTN reservoir statistics, in contrast to the MPS-RC case where the corresponding thermodynamic-limit transition point is at $\sigma_T=0$. This analogy should be understood at the level of mean-field reservoir statistics rather than as a claim of thermodynamic equivalence. Related intuition can be drawn from the presence or absence of finite-temperature phase transitions in Ising-type models, where topology, long-range interactions, or effective dimensionality can change the transition behavior \cite{baxter2007exactly,goldenfeld1992lectures}.

A scaling analysis can be performed for the mean-field order parameter near criticality. Under the assumption that $0 \le \epsilon \ll 1$ where $\epsilon\equiv(\tilde{N}_x-1)(\frac{2-\sigma_T^2}{2})$, the scaling hypothesis is $\left\langle x_\gamma(t)^2 \right\rangle  \approx a-b\epsilon $, where $a$ and $b$ are numerical constants. The scaling function does not depend on the input distribution, the input dimension, and the bond dimension. Details and justification of the calculations are provided in \ref{subsec:scaling_apndx}. The critical exponent with respect to $\tilde{N}_x \sigma_T^2$ is similar to the MPS-RC case under a different encoding, but the scaling with respect to the bond dimension is different. These results are different from, yet might be related to, the universal behaviors studied in TN generative models and random pair models \cite{Bai2024,Lu2025}. Further investigation is needed to determine whether this similarity is superficial or reflects a more general universality phenomenon.

Measurement-induced entanglement phase transitions in random quantum circuits are believed to be related to phase transitions in random tree tensor networks (RTTNs) \cite{Li2018QuantumZenoEffect,Skinner2019MeasurementInducedPhase,Chan2019UnitaryProjective,Jian2020MeasurementCritic,lopez2020TransitionTTN}. Since RC has been proposed as a method for probing phase transitions in quantum-circuit reservoirs \cite{kobayashiFeedbackDrivenQuantumReservoir2024,kobayashiQuantumReservoirProbing2025a}, it is also instructive to compare our work with a previous mean-field study of entanglement phase transitions in RTTNs \cite{lopez2020TransitionTTN}.
The random TTN elements in Ref.~\cite{lopez2020TransitionTTN} are drawn from a zero-mean unit-variance Gaussian distribution. If we translate the setting of Ref.~\cite{lopez2020TransitionTTN} into our setting (coordination number $q=3$ and variance $\sigma_T^2/\chi^2$), then the entanglement critical point is given by $\sigma_T^{ent}=e^{2\tanh^{-1}(1/2) }=3$, which is also a constant with respect to the bond dimension, but the value is different from $\sqrt{2}$. This suggests a similar scaling behavior, but the exact locations of critical lines are different. It would be interesting to study their potential links.

\section{Conclusion}
\label{sec:conclusion}

In this work, we introduced Tree Tensor Network Reservoir Computing (TTN-RC). In the tested NARMA tasks, TTN-RC demonstrated competitive or improved performance compared with the standard ESN baseline, especially in higher-order settings. We established theoretical tools for model selection, including an expected contraction indicator $C$ for stability analysis and a mean-field theory for reservoir-state statistics.

Furthermore, we identified an asymptotic phase-boundary-like behavior at $\sigma_T = \sqrt{2}$ in the thermodynamic limit for multiple theoretical indicators. This research demonstrates the potential of tensor-network-based reservoir computing and advances our understanding of critical phenomena within reservoir computing. Furthermore, it suggests a route for applying metrics traditionally used in the analysis of classical reservoir computing, as well as phenomenological tools from learning theory, to quantum-inspired and future quantum reservoir computing models. This research presents a theoretically tractable RC model and provides new insights into how hierarchical tensor-network topology can control the stability and computational capability of reservoir systems.

Although this study focused on a comparison between TTN-RC and the ESN baseline,
more performance evaluations against the MPS-RC on a variety of tasks would be desirable. The tensor networks employed in this study were limited to being quantum-inspired and did not carry the direct meaning of quantum operations.
\enlargethispage{2\baselineskip}
Looking ahead, we aim to develop tensor networks as quantum channels for more controllable digital quantum reservoir computing and stochastic reservoir computing \cite{ehlersStochasticReservoirComputers2025}. Their connectivity and information-scrambling properties will be an important direction for future research.

\newpage
\appendix

\section{Jacobian-based Echo State Property}
\label{subsec:ESP}
Echo State Property (ESP) is one of the core characteristics of a reservoir computing system, meaning that the system's internal state is uniquely determined solely by the history of external inputs \cite{jaegerEchoStateApproach, buehnerTighterBoundEcho2006, yildizRevisitingEchoState2012, gallicchioChasingEchoState2019}. Concretely, after a sufficiently long time, the reservoir state completely loses any dependence on its initial conditions and depends only on the past input sequence. This phenomenon is the condition for the system to be stable and convergent.

By optimizing the reservoir's hyperparameters to satisfy the ESP, reservoir computing can achieve stable training and reliable predictions.

For reservoir states $\bm{x}(t)$ and $\bm{x}'(t)$ evolving from different initial reservoir states $\bm{x}(t_0)$ and $\bm{x}'(t_0)$, the ESP is satisfied if
\begin{equation}
    \lim_{t \rightarrow \infty} \|\bm{x}(t) - \bm{x}'(t)\| \rightarrow 0
\end{equation}
holds, where $\|\cdot\|$ denotes the Euclidean norm.

In standard ESNs, the ESP can be ensured by adjusting the spectral radius of the reservoir weight matrix, based on the condition that each step of the reservoir acts as a contraction mapping \cite{jaegerEchoStateApproach}.

However, in our TTN-RC, since we utilize random tensor networks as reservoirs, the equivalent concept to the spectral radius is not immediately obvious. Therefore, we use an expected contraction condition derived from expectation operations and the Jacobian of the reservoir as an analytical stability indicator. This method has also been successfully applied in research on random MPS reservoirs \cite{Sato2025}.

Let $\tilde{N}_x=N_x/M$ and $\tilde{L}=\log_2\tilde{N}_x$ denote the size and depth of one tree; for $M=1$, these reduce to $\tilde{N}_x=N_x$ and $\tilde{L}=L$. We write the per-tree reservoir state update as $r_\gamma(\bm{x}, \bm{u}):=f(g_\gamma^{(\tilde{L})} (\bm{x}) + \sum_{k=1}^{N_u} W_{\gamma k}^\text{(in)} u_k)$ for $\gamma=1,\ldots,\tilde{N}_x$.
For instance, making the time explicit, this state update function $\bm{r}$ takes the reservoir state at the previous time step, $\bm{x}(t-1)$, and the external input signal at the current time step, $\bm{u}(t)$, as arguments and returns the reservoir state at the current time step.
$f$ is the activation function and $W^\text{(in)} \in \mathbb{R}^{\tilde{N}_x\times N_u}$ is the random weight function for encoding of external inputs within one tree. Also, $g_\gamma^{(\tilde{L})} (\bm{x})$ is the result of transforming $\bm{x}$ with the random tensor network.

Suppose this state update function $\bm{r}$ possesses randomness, similar to the elements of a tensor. In this case, for two different reservoir states $\bm{x}, \bm{x}'$ and any fixed input $\bm{u}$, the following holds:
\begin{equation}
    \left\langle \|\bm{r}(\bm{x}, \bm{u}) - \bm{r}(\bm{x}', \bm{u})\|\right\rangle_T \leq C \|\bm{x}-\bm{x}'\| \ .
\end{equation}
Here, $C:=\tilde{N}_x K \Lambda$, where $K$ is the Lipschitz constant of the activation function included in $\bm{r}$. In TTN-RC, sigmoid function is used, so $K=1/4$. Furthermore, $\Lambda$ is a quantity related to the Jacobian of $g_\gamma^{(\tilde{L})} (\bm{x})$ with respect to $\bm{x}$, satisfying $\sqrt{\langle (\frac{\partial g_{\gamma}^{(\tilde{L})}(\bm{x})}{\partial x_{m}})^2 \rangle_T} \leq \Lambda$ for all reservoir states $\bm{x} \in [0,1]^{\tilde{N}_x}$.
Fortunately for our calculations, in TTN-RC, $\langle (\frac{\partial g_{\gamma}^{(\tilde{L})}(\bm{x})}{\partial x_{m}})^2 \rangle_T$ is independent of $\bm{x}$. Therefore, $\Lambda = \sqrt{\langle (\frac{\partial g_{\gamma}^{(\tilde{L})}(\bm{x})}{\partial x_{m}})^2 \rangle_T}$.

$C<1$ indicates that the reservoir state update is a contraction mapping in expectation under the assumptions above. This expected-contraction regime is used as an analytical proxy for stable dynamics in which the reservoir state loses dependence on its initial condition. Conversely, when $C \ge 1$, this expectation-level condition is not satisfied, and the reservoir's dynamics can become sensitive to initial conditions. Based on this discussion, we use $C=1$ as a candidate analytical boundary between stable and sensitive regimes.

 \section{Mean-field learning theoretical phenomenology}
\label{subsec:GGO}

The Gonon-Grigoryeva-Ortega (GGO) RC generalization theory is used \cite{gononRiskBoundsReservoir2020}. The definitions of the quantities and the assumptions can be found in \cite{gononRiskBoundsReservoir2020}. The generalization gap upper bound (from Theorem~14 of \cite{gononRiskBoundsReservoir2020}) for the RC hypothesis set $H^{RC}$ is $\mathbb{P}[\sup_{H\in H^{RC}}|\hat R_{n_{data}}(H)-R(H)| \le \Delta_g]\ge 1-\delta$ over the randomness of input and target stochastic processes $(Z_{-i},Y_{-i})_{i \in \{0,1,...,n_{data}-1\}}$, where $R(H)=\mathbb{E} L(H(Z),Y_0)$ and $\hat R_{n_{data}}=\frac{1}{n_{data}} \sum_{i=0}^{n_{data}-1} L(H(Z_{-i}^{-n_{data}+1}),Y_{-i})$ for some loss function $L$, and $n_{data}$ is the number of training data steps. The bound $\Delta_g$ is
\begin{align} \label{eq:ggo}
\Delta_g = (C_0+C_1)\frac{1}{n_{data}}+ C_2 \frac{\log(n_{data})}{n_{data}} +C_3 \sqrt{\frac{\log(n_{data})}{n_{data}}} +C_{bd}\sqrt{\frac{\log (4/\delta)}{2n_{data}}} ,
\end{align}
where the parameters are listed in Table~\ref{tab:GGOparameters}. $\rho\in (0,1)$ is the contraction constant of $\bm{r}(\bm{x},\bm{u})$ with respect to the reservoir state $\bm{x}$. The mean-field value of $\rho$ is used for the evaluation, and its calculation is given in \ref{subsec:ESPindex}. The output dimension is assumed to be $N_y=1$, so $\Lambda_{W_{out}}$ in \cite{Sato2025} equals $\overline{L_h}$ in \cite{gononRiskBoundsReservoir2020}. $C_{RC}$ is defined through a GGO Rademacher complexity upper bound $\mathcal R_{n_{data}}\le \frac{C_{RC}}{\sqrt{n_{data}}} $ (Corollary~8 of \cite{gononRiskBoundsReservoir2020}). In our calculation, the expected GGO Rademacher complexity is upper bounded by the expected kernel of the reservoir feature map $\mathbb{E}_T \mathcal R_{n_{data}}\le \frac{ \Lambda_{W_{out}} }{\sqrt{n_{data}}} \sqrt{\langle x^2_\gamma \rangle+ \max_{\bm{v}_{in}} [2K\sqrt{N_x}\lVert \bm{v}_{in} \rVert +(K\lVert \bm{v}_{in} \rVert)^2] }$ using Lemma~3.9 of \cite{Sato2025}, where the expectation is over the randomness of the reservoir and $\bm v_{in}\equiv W_{in} \bm u(t)$. In our experiments, $\lVert \bm{v}_{in} \rVert \le \sqrt{N_x}$, so $C_{RC}=\sqrt{ \langle x^2_\gamma \rangle+K(2+K)N_x}$ under the assumption that $\Lambda_{W_{out}}= 1$. Since $Y$ is the unknown target process, the unknown quantities of $Y$ are assumed to be equal to their counterparts of $Z$. For example, we assume the weight sequences $w^y_j=w^z_j=2^{-j}$. The bound diverges when $\rho \rightarrow 1$, so the truncated value $\bar \rho=\min \{0.999, \rho \}$ is used for the numerical evaluation of $\Delta_g$ in Fig.~\ref{fig:theory}(d). The confidence probability is $\delta=0.4$. The loss function is the square loss. The GGO theory predicts a second-descent phenomenon \cite{Belkin2019,Nakkiran2021} with respect to $\sigma_T$ through $\rho$, but whether this behavior depends on other parameters remains to be investigated.

\begin{table}[t]
\begin{center}
\resizebox{\linewidth}{!}{%
\begin{tabular}{ c c c c c }
 \hline
 GGO & TNRC & Value & Ref. \cite{gononRiskBoundsReservoir2020} & Ref. \cite{Sato2025} \\
 \hline
  $m$ & $N_y$ & 1 &  P. 5 &  N/A \\
  $d$ & $N_u$ & 1 & P. 5 & N/A \\
  $D_{w^I}, I\in\{y,z\}$ & N/A & 1/2 & P. 6 & N/A \\
  $L_{w^I}, I\in\{y,z\} $ & N/A & 2 & P. 6 & N/A \\
  $N$ & $N_x$ & 256 & P. 7 &  N/A \\
  $n$ & $n_{data}$ & $ 2\times10^{5}$ & P. 13 & N/A \\
  $L_I, I\in\{y,z\}$ & upper bound of $\lVert \bm{v}_{in} \rVert$ & $\sqrt{N_x}$ & P. 15, (24) & N/A \\
  $L_R$ & $K$ & 1/4 & P. 18, Assumption 4 & Lemma 3.1 \\
  $M_F$ & upper bound of $|y(t)|$ & 1   & P. 18, Assumption 6 & N/A \\
  $L_L$ &  upper bound of $2|y(t)|$  & 2 & P. 20 & N/A \\
  $L_{h,0}=\overline L_h$ & $\Lambda_{W_{out}}$  & 1 &  P. 19-20 & Lemma 3.9 \\
  $\mathbb{E}[L(\bm 0,\bm Y_0)]$ & N/A & 1 & P. 24, (44) & N/A \\
  $\mathbb{E} \lVert \bm \xi_0^I \rVert_2, I\in\{y,z\}$ & $\mathbb{E} |u(0)|$ & 1/2 & P. 25, Corollary 8 & N/A \\
  $\overline M$ & upper bound of $|u(t)|$ & 1 & P. 45, Proposition 19  & N/A\\
 \hline
\end{tabular}
}
\end{center}

\begin{center}
\resizebox{\linewidth}{!}{%
\begin{tabular}{ c c c c }
 \hline
 GGO & Value & Ref. \cite{gononRiskBoundsReservoir2020} \\
 \hline
  $C_0=2 L_L\overline L_h M_F \frac{\rho}{1-\rho}$ &  $ 4 \frac{\rho}{1-\rho}$  & P. 21, (39)  \\
  $M=L_L (\overline L_h M_F + L_{h,0})+\mathbb{E}[L(\bm 0,\bm Y_0)]$ & 5  &   P. 24, (44) \\
  $\lambda_{max}=\max (\rho,D_{w^I}), I\in\{y,z\}$ & $\max (\rho,1/2)$ & P. 25, Corollary 8\\
  $C_I= \frac{2L_I \mathbb{E}\lVert \bm \xi_0^I \rVert_2}{1-D_{w^I}} , I\in\{y,z\}$ & $2\sqrt{N_x}$ & P. 25, Corollary 8 \\
  $C_1=\frac{2 L_L\overline L_h M_F+L_L C_y}{\lambda_{max}}$ & $\frac{4+2\sqrt{N_x}}{\lambda_{max}}$ &  P. 25, (51) \\
  $C_2=\frac{2M}{\log (\lambda_{max}^{-1})}+ \frac{L_L L_R \overline L_h C_z}{\lambda_{max}\log (\lambda_{max}^{-1})}$ & $\frac{10}{\log (\lambda_{max}^{-1})}+ \frac{\sqrt{N_x}}{\lambda_{max}\log (\lambda_{max}^{-1})}$ &  P. 25, (51) \\
  $C_3=\frac{2\sqrt{m}L_L C_{RC}}{\sqrt{\log (\lambda_{max}^{-1})}}$ & $4\sqrt {\frac{  \langle x^2_\gamma \rangle+9N_x/16}{{\log (\lambda_{max}^{-1})}}}$ &  P. 25, (52) \\
  $C_{bd}\ \text{(see caption)}$ & $4(\frac{1}{1-\rho}(\rho+\frac{\sqrt{N_x}}{2})+2\sqrt{N_x})$ &  P. 45, (93) \\
 \hline
\end{tabular}
}
\end{center}
\caption{Parameters used for Eq.~\ref{eq:ggo} to calculate Fig.~\ref{fig:theory}(d). The GGO column is the expression from Ref.~\cite{gononRiskBoundsReservoir2020}. The TNRC column is the corresponding quantity in the current manuscript or in Ref.~\cite{Sato2025}. The Value column is the formula used in the numerical evaluation. The locations in references are provided when possible. $C_{bd}=2L_L(\frac{\overline L_h}{1-\rho}(M_F\rho+L_R\overline M L_z \lVert w^z \rVert_1)+\overline M L_y \lVert w^y\rVert_1) $.}
\label{tab:GGOparameters}
\end{table}

\FloatBarrier
\section{Experimental Generalization Performance}
\label{subsec:generalize}

\begin{figure*}[t]
\centering
\includegraphics[width=\linewidth]{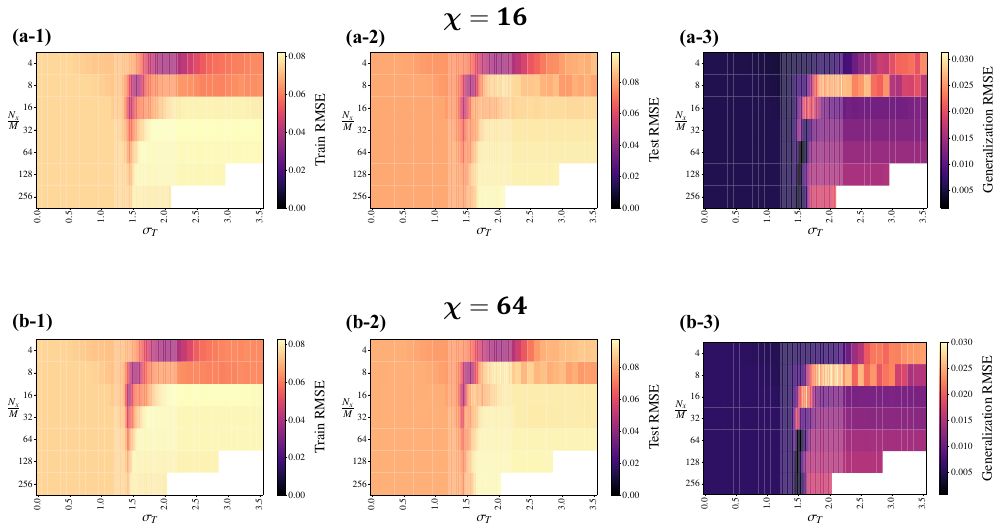}
\caption{Heatmaps of RMSE on the training and test data, and the generalization performance for the NARMA5 task. The experimental settings are the same as in Fig.~\ref{fig:mc_narma_heatmap}.}
\label{fig:NARMA5_gene}
\end{figure*}

Here, to investigate the parameter dependence of the generalization performance on the NARMA5 task, we show heatmaps of the training and test performance, corresponding to Fig.~\ref{fig:mc_narma_heatmap}(b). We use the root mean squared error (RMSE) as the performance metric:
\begin{equation}
\text{RMSE} = \sqrt{\frac{1}{T} \sum_{t=1}^{T} (d(t) - y(t))^2}
\end{equation}
Figure~\ref{fig:NARMA5_gene} shows the heatmaps of the RMSE for the training data, test data, and the generalization error. The generalization RMSE is defined as
\begin{equation}
\text{Generalization RMSE} = \text{Test RMSE} - \text{Train RMSE}
\end{equation}
Although the trends of the Train RMSE and Test RMSE appear nearly identical, the Train RMSE consistently shows lower values. It is observed that the heatmap for the generalization RMSE, which is the difference between them, resembles the plot in Fig.~\ref{fig:mc_narma_heatmap}(d). Generalization RMSE can be large in regions where the expected-stability indicator does not support contraction.

\FloatBarrier
\section{Scaling analysis near criticality}
\label{subsec:scaling_apndx}

\begin{figure*}[t]
    \centering
    \includegraphics[width=0.99\linewidth]{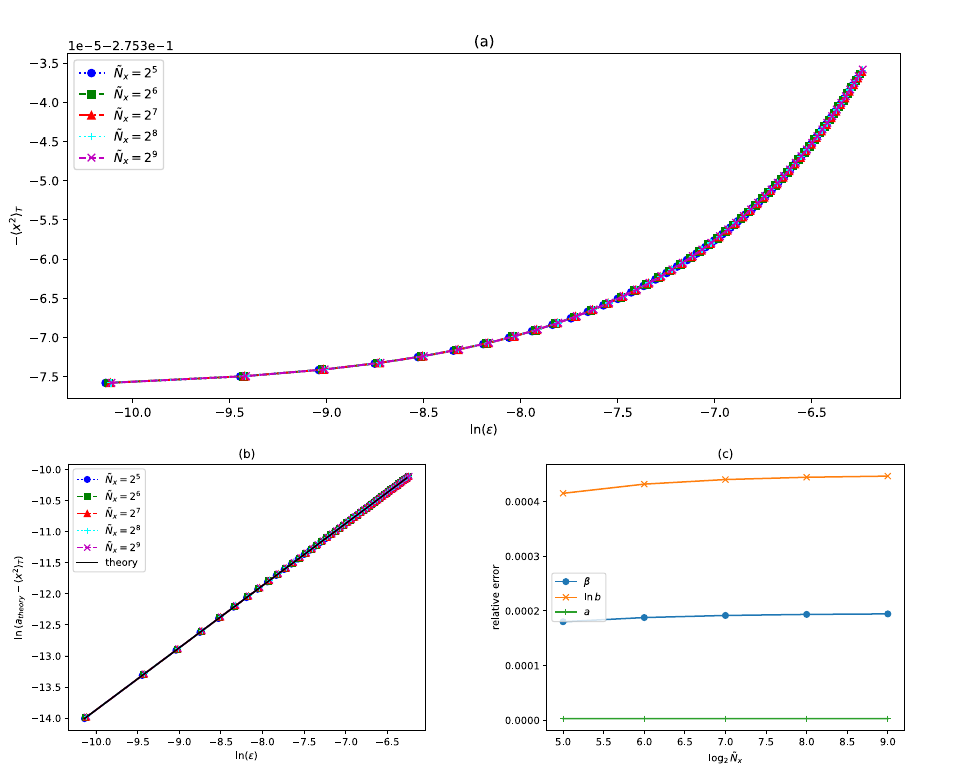}
    \caption{Scaling behavior near the critical point. (a) Mean-field data near the critical point. (b) Shifted data on a logarithmic scale compared with the theory. (c) Relative error of the estimated parameters.}
    \label{fig:scaling}
\end{figure*}

The mean-field order parameter near the critical point is plotted in Fig.~\ref{fig:scaling}(a), where universal scaling behavior can be observed. The scaling hypothesis is $a-\langle x_\gamma^2 \rangle = b \epsilon^\beta $ for $\epsilon\equiv(\tilde{N}_x-1)(\frac{2-\sigma_T^2}{2})$ from the theory. The parameters $(\ln{b},\beta)$ are estimated through linear regression of the shifted data on a logarithmic scale in Fig.~\ref{fig:scaling}(b). The parameter $a$ can be estimated from the maximum of the data. These estimates are compared with the theoretical values $a_{theory}=\frac{1}{4}+\frac{1}{4}\int \mathcal{D}z [\tanh^2(k)]$, $b_{theory}=\frac{1}{4}\int \mathcal{D}z [\frac{k\tanh(k)}{\cosh^2(k)}]$, and $\beta_{theory}=1$, where $k\equiv \frac{z}{2\sqrt{2}}$ and $\mathcal{D}z\equiv \frac{1}{\sqrt{2\pi}}e^{-z^2/2} dz$. The relative error is plotted in Fig.~\ref{fig:scaling}(c). The relative error does not change significantly with respect to $\tilde{N}_x$. The theoretical expansion to the lowest power in $\epsilon$ is
\begin{align}
V_g = & \frac{1}{2}\left(1-\frac{2-\sigma_T^2}{2}\right)^{\tilde{N}_x-1} & \\
\approx &   \frac{1}{2} (1-\epsilon) & \\
\langle x_\gamma^2 \rangle = & \frac{1}{4}\left[1+\int \mathcal{D}z \tanh^2\left(\frac{\sqrt{V_g}z}{2}\right)\right] & \\
 \approx & \frac{1}{4}+\frac{1}{4}\int \mathcal{D}z \left[\tanh^2(k)-\frac{k\tanh(k)}{\cosh^2(k)} \epsilon\right]. &
\end{align}

\FloatBarrier
\section{Average ESP for multiple different initial state pairs and mean-field contraction constant}
\label{subsec:ESPindex}

\begin{figure*}[b]
\centering
\includegraphics[width=\linewidth]{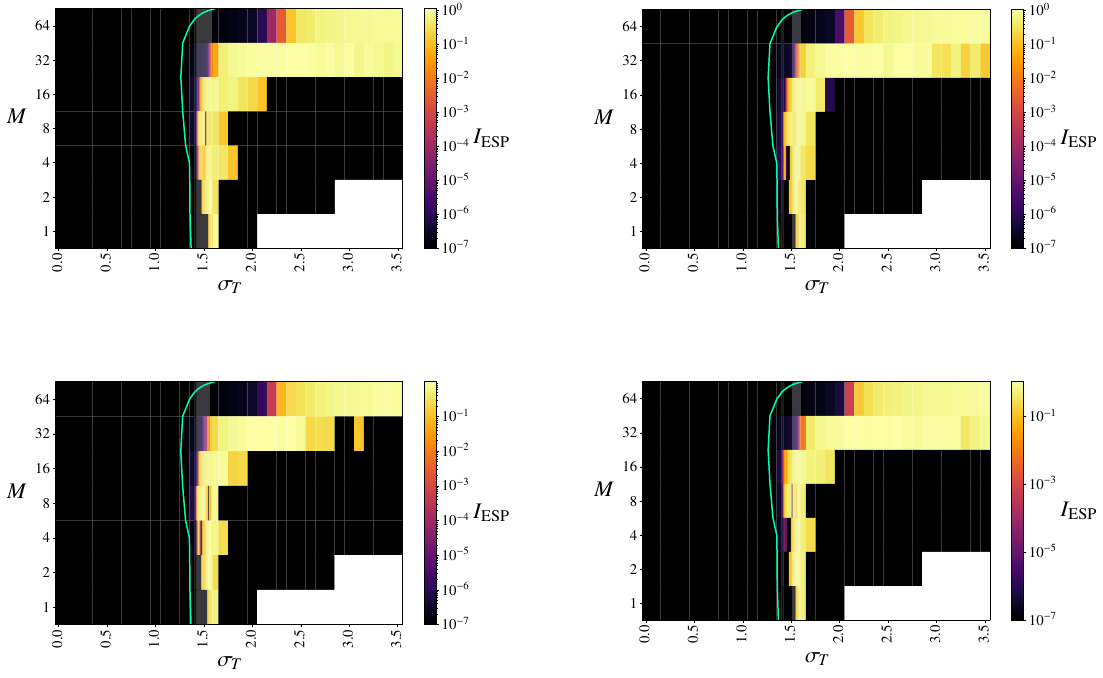}
\caption{Heatmap of the ESP metric, $I_{\text{ESP}}(100)$. The light-green line indicates the contour for $C=1$. In panel (d), values of $10^{-7}$ or less are displayed in the same color.
Values of $I_\text{ESP}(100)$ are averaged over 10 different initial-state pairs.
The four heatmaps correspond to four different reservoir realizations.
The white regions in the heatmaps indicate parameter settings where the TTN computation diverged, making subsequent calculations impossible.
}
\label{fig:esp_diff_ini}
\end{figure*}

Fig.~\ref{fig:mc_narma_heatmap}(d) presented a heatmap of the mean $I_\text{ESP}$ averaged over 10 different reservoir realizations for a fixed pair of initial states. The results are consistent with the expected stability indicator $C < 1$, and a recovery of the small-ESP-index regime occurs as $\sigma_T$ increases. Furthermore, we observed that the task performance is maximized in a region where $C > 1$, showing that the expected-contraction indicator alone is not sufficient for identifying optimal task performance.
Here, in Fig.~\ref{fig:esp_diff_ini}, we show heatmaps of the mean $I_\text{ESP}$ averaged over 10 different pairs of initial states for a fixed reservoir realization. The results for four different reservoir realizations are presented. These results are strikingly similar to those in Fig.~\ref{fig:mc_narma_heatmap}(d) and are consistent with the interpretation that $C<1$ provides an expected sufficient contraction condition. Furthermore, the phenomenon in which the small-ESP-index region appears twice is also observed.

\begingroup

In order to understand $I_\text{ESP}$, a mean-field calculation for the Lipschitz constant is performed \cite{toyoizumiEdgeChaosAmplification2011,massarMeanfieldTheoryEcho2013}. Consider two non-random points $\bm{x}'(t)$ and $\bm{x}(t)$ that are close to each other. The random variables of the iteration are approximately
\begin{align*}
     x'_i(t+1)-x_i(t+1)  &= f(a_i'(t))-f(a_i(t)) \\
    & \approx f'\!\left(\frac{a_i'(t)+a_i(t)}{2}\right) (a_i'(t)-a_i(t)) ,
\end{align*}
\begin{align} \label{eq:aa}
      a_i'(t)-a_i(t) & = g_i(\bm{x}'(t)) - g_i(\bm{x}(t)) \\
     &\approx  \sum_j \left.\frac{\partial g_i}{\partial x_j}\right\vert_{\frac{\bm{x}'(t)+\bm{x}(t)}{2}}(x'_j(t)-x_j(t)) . \notag
\end{align}
This is a stochastic difference equation. Under the assumption that $\{\Psi^\gamma_{i_1\cdots i_{\tilde{N}_x}}\}$ is jointly normally distributed, $(a_i'(t),a_i(t))$ is jointly normally distributed, and hence $(a_i'(t)+a_i(t),a_i'(t)-a_i(t))$ is jointly normally distributed. Since $\text{Cov}(a_i'(t)+a_i(t),a_i'(t)-a_i(t))=\text{Var}(a_i'(t))-\text{Var}(a_i(t))=0$, $a_i'(t)+a_i(t)$ and $a_i'(t)-a_i(t)$ are uncorrelated jointly normally distributed random variables, and hence are independent \cite{Dudley_2002}. The mean square is then approximately
\begin{align*}
    & \langle [x'_i(t+1)-x_i(t+1)]^2 \rangle_T
    \approx
    \left\langle \left[f'\!\left(\frac{a_i'(t)+a_i(t)}{2}\right)\right]^2 \right\rangle_T
    \langle [a_i'(t)-a_i(t)]^2 \rangle_T .
\end{align*}
Using zero cross-covariance between different input coordinates, the last mean square is
\begin{align*}
     \langle [a_i'(t)-a_i(t)]^2 \rangle_T  \approx V_J \|\bm{x}'(t)-\bm{x}(t)\|^2 ,
\end{align*}
where, for $\mu \ne \nu$,
\begin{align*}
    \left\langle \frac{\partial g_\gamma}{\partial x_\mu}\frac{\partial g_\gamma}{\partial x_\nu} \right\rangle_T
    &= \sum_{\{i_k\},\{j_k\}}
    \left\langle \Psi^\gamma_{i_1\cdots i_{\tilde{N}_x}} \Psi^\gamma_{j_1\cdots j_{\tilde{N}_x}} \right\rangle_T \\
    &\quad \times
    \phi'_{i_\mu}(x_\mu)\phi_{j_\mu}(x_\mu)
    \phi_{i_\nu}(x_\nu)\phi'_{j_\nu}(x_\nu)
    \prod_{k\notin \{\mu,\nu\}} \phi_{i_k}(x_k)\phi_{j_k}(x_k) \\
    &=0 .
\end{align*}
Summing over the output coordinate $i$ gives
\begin{align*}
 \frac{\left\langle \|\bm{x}'(t+1)-\bm{x}(t+1)\|^2 \right\rangle_T}{\|\bm{x}'(t)-\bm{x}(t)\|^2}
 \approx \tilde{N}_x V_J
 \left\langle \left[f'\!\left(\frac{a_i'(t)+a_i(t)}{2}\right)\right]^2 \right\rangle_T ,
\end{align*}
which can be interpreted as a mean-field calculation for $I_{\text{ESP}}$. When the mean-field Lipschitz constant
\begin{align*}
\rho=\sqrt{\tilde{N}_x V_J
\left\langle \left[f'\!\left(\frac{a_i'(t)+a_i(t)}{2}\right)\right]^2 \right\rangle_T}
\end{align*}
is smaller than one, this calculation indicates contraction in mean square with contraction constant $\rho$. When $\rho$ is larger than one, the mean-field contraction condition is not satisfied, and the reservoir can enter a high-sensitivity regime. In Fig.~\ref{fig:theory}(c), $\rho$ is plotted.
\endgroup

\FloatBarrier
\section{Details of Calculations}

\label{subsec:calculations}

\subsection{Statistical Properties of TTN Function}
\label{subsec:ttn_stat}
We seek the expectation with respect to the randomness of $T$, denoted as $\left\langle\cdot\right\rangle_{T} := \mathbb{E}_{p(T)}[\cdot]$.
Throughout this appendix, we use $n_l=2^l$ and $n_{l-1}=n_l/2$, consistently with the notation in the main text. For the top layer of one tree, $n_{\tilde{L}}=\tilde{N}_x$.
First, it can be transformed as follows:
\begingroup

\begin{equation}
\begin{split}
g_{\gamma}^{(l)}(\bm{x}) &= \sum_{I(i_{1}:i_{n_l})=(0,0,...,0)}^{(1,1,...,1)} \Psi_{I(i_{1}:i_{n_l})}^{\gamma,(l,1)} \prod_{n=1}^{n_l} \phi_{i_{n}}(x_{n}) \\
&= \sum_{\alpha,\beta} T_{\alpha,\beta}^{\gamma,(l,1)} \sum_{I(i_{1}:i_{n_l})=(0,0,...,0)}^{(1,1,...,1)} \Psi_{I(i_{1}:i_{n_{l-1}})}^{\alpha,(l-1,1)} \Psi_{I(i_{n_{l-1}+1}:i_{n_l})}^{\beta,(l-1,2)} \prod_{n=1}^{n_l} \phi_{i_{n}}(x_{n}) \\
&= \sum_{\alpha,\beta} T_{\alpha,\beta}^{\gamma,(l,1)} g_{\alpha}^{(l-1)}(x_{1},...,x_{n_{l-1}}) g_{\beta}^{(l-1)}(x_{n_{l-1}+1},...,x_{n_l}) .
\end{split}
\end{equation}
\endgroup

Therefore,
\begin{equation}
\left\langle g_{\gamma}^{(l)}(\bm{x})\right\rangle_{T}
= \sum_{\alpha,\beta} \left\langle T_{\alpha,\beta}^{\gamma,(l,1)}\right\rangle_{T} \left\langle g_{\alpha}^{(l-1)}(x_{1},...,x_{n_{l-1}})\right\rangle_{T} \left\langle g_{\beta}^{(l-1)}(x_{n_{l-1}+1},...,x_{n_l})\right\rangle_{T} = 0
\end{equation}

can be calculated.
Here, note that $g_{\alpha}^{(l-1)}(x_{1},...,x_{n_{l-1}})$ and $g_{\beta}^{(l-1)}(x_{n_{l-1}+1},...,x_{n_l})$ are independent in the ensemble of $T$.

Next, we calculate the covariance $C_{\eta\theta}^{(l)} := \left\langle g_{\eta}^{(l)}g_{\theta}^{(l)}\right\rangle_{T}$.

\begin{equation}
\begin{split}
C_{\eta\theta}^{(l)}(\bm{x})
&= \left\langle g_{\eta}^{(l)}(\bm{x})g_{\theta}^{(l)}(\bm{x})\right\rangle_{T} \\
&= \left\langle \left(\sum_{\alpha,\beta}T_{\alpha,\beta}^{\eta,(l,1)}{g_{\alpha}^{(l-1)}(x_{1},...,x_{n_{l-1}})g_{\beta}^{(l-1)}(x_{n_{l-1}+1},...,x_{n_l})}\right) \right. \\
& \quad \left. \times \left(\sum_{\alpha^{\prime},\beta^{\prime}}T_{\alpha^{\prime},\beta^{\prime}}^{\theta,(l,1)}{g_{\alpha^{\prime}}^{(l-1)}(x_{1},...,x_{n_{l-1}})g_{\beta^{\prime}}^{(l-1)}(x_{n_{l-1}+1},...,x_{n_l})}\right) \right\rangle_{T} \\
&= \left\langle \sum_{\alpha,\beta,\alpha',\beta'} (T_{\alpha,\beta}^{\eta,(l,1)}T_{\alpha',\beta'}^{\theta,(l,1)}) (g_{\alpha}^{(l-1)}g_{\alpha'}^{(l-1)})(g_{\beta}^{(l-1)}g_{\beta'}^{(l-1)}) \right\rangle_{T} \\
&= \sum_{\alpha,\beta} \left\langle T_{\alpha,\beta}^{\eta,(l,1)}T_{\alpha,\beta}^{\theta,(l,1)}\right\rangle_{T} {\left\langle(g_{\alpha}^{(l-1)}(x_{1},...,x_{n_{l-1}}))^{2}\right\rangle_{T} \left\langle(g_{\beta}^{(l-1)}(x_{n_{l-1}+1},...,x_{n_l}))^{2}\right\rangle_{T}} \\
&= \delta_{\eta\theta} \sum_{\alpha,\beta}\frac{\sigma_T^{2}}{{d_\alpha d_\beta}} {C_{\alpha\alpha}^{(l-1)}(x_{1},...,x_{n_{l-1}})C_{\beta\beta}^{(l-1)}(x_{n_{l-1}+1},...,x_{n_l})}
\end{split}
\end{equation}
From the calculation above, we can see that the off-diagonal elements are zero.

Let the diagonal components of the covariance matrix, i.e., the variance, be $V_g^{(l)} (\bm{x}) := C_{\alpha\alpha}^{(l)}(\bm{x})$ .
This variance does not depend on the bond dimension indices $\alpha$ and $\beta$ (which is evident from the calculation for the $l=2$ case).

Thus, we obtain
\begingroup

\begin{align}
C_{\alpha \alpha}^{(l-1)}(x_1, \cdots, x_{n_{l-1}})
&= V_g^{(l-1)}(x_1, \cdots, x_{n_{l-1}}) \\
C_{\beta \beta}^{(l-1)}(x_{{n_{l-1}}+1}, \cdots, x_{n_l})
&= V_g^{(l-1)}(x_{{n_{l-1}}+1}, \cdots, x_{n_l})
\end{align}
\endgroup

and obtain the following asymptotic formula for the variance:
\begingroup

\begin{equation}
\begin{split}
    &V_g^{(l)}(x_1, \cdots, x_{n_l}) \\
    &= \sigma_T^2 V_g^{(l-1)}(x_1, \cdots, x_{n_{l-1}}) V_g^{(l-1)}(x_{{n_{l-1}}+1}, \cdots, x_{n_l}) \ .
\end{split}
\end{equation}
\endgroup

Next, we calculate $V_g^{(1)}(x_1,x_2)$ as follows:
\begin{equation}
\begin{split}
V_{g}^{(1)}(x_{1},x_{2})
&= \left\langle \left(\sum_{i_{1},i_{2}}T_{i_{1},i_{2}}^{\gamma,(1,1)}\phi_{i_{1}}(x_{1})\phi_{i_{2}}(x_{2})\right) \left(\sum_{i_{1}^{\prime},i_{2}^{\prime}}T_{i_{1}^{\prime},i_{2}^{\prime}}^{\gamma,(1,1)}\phi_{i_{1}^{\prime}}(x_{1})\phi_{i_{2}^{\prime}}(x_{2})\right) \right\rangle_{T} \\
&= \sum_{i_{1},i_{2}} \left\langle(T_{i_{1},i_{2}}^{\gamma,(1,1)})^{2}\right\rangle_{T} (\phi_{i_{1}}(x_{1}))^{2}(\phi_{i_{2}}(x_{2}))^{2} \\
&= \frac{\sigma_T^2}{{2\cdot 2}} \\
&= \frac{\sigma_T^2}{4} \ .
\end{split}
\end{equation}

Therefore, $V_{g}^{(l)}(x_{1},...,x_{n_l})$ does not depend on $\bm{x}$,
\begin{equation}
    V_{g}^{(l)} = \sigma_T^{2}(V_{g}^{(l-1)})^{2}
\end{equation}
is obtained.

Solving this recurrence relation, we calculate the variance of the TTN function to be:
\begin{equation}
    V_{g}^{(l)} = \frac{(\sigma_T^2)^{n_l-1}}{2^{n_l}} \ .
\end{equation}

For the top layer of one tree, $l=\tilde{L}$ and $n_{\tilde{L}}=\tilde{N}_x$; for $M=1$, this reduces to $\tilde{N}_x=N_x$. Thus, the variance can be written as:
\begin{equation}
V_{g}^{(\tilde{L})} = \frac{(\sigma_T^2)^{n_{\tilde{L}}-1}}{2^{n_{\tilde{L}}}} = \frac{(\sigma_T^2)^{\tilde{N}_x-1}}{2^{\tilde{N}_x}} \ .
\end{equation}

\subsection{Statistical Properties of the Reservoir Jacobian Matrix}
\label{subsec:jacobi_stat}
Consider the Jacobian matrix of the random tree tensor network function. Assume $1 \le m \le n_{l-1}$ in the process.
\begingroup

\begin{equation}
\begin{split}
J_{\gamma m}^{(l)}(\bm{x}) &:= \frac{\partial g_{\gamma}^{(l)}(\bm{x})}{\partial x_{m}} \\
&= \sum_{I(i_{1}:i_{n_l})=(0,0,...,0)}^{(1,1,...,1)} \Psi_{I(i_{1}:i_{n_l})}^{\gamma,(l,1)} \frac{\partial\phi_{i_{m}}(x_{m})}{\partial x_{m}} \prod_{n=1,n\ne m}^{n_l} \phi_{i_{n}}(x_{n}) \\
&= \sum_{I(i_{1}:i_{n_l})=(0,0,...,0)}^{(1,1,...,1)} \Psi_{I(i_{1}:i_{n_l})}^{\gamma,(l,1)} \frac{\partial\phi_{i_{m}}(x_{m})}{\partial x_{m}} \frac{1}{\phi_{i_{m}}(x_{m})} \prod_{n=1}^{n_l} \phi_{i_{n}}(x_{n}) \\
&= \sum_{\alpha,\beta} T_{\alpha,\beta}^{\gamma,(l,1)} \left(\sum_{I(i_{1}:i_{n_{l-1}})} \Psi_{I(i_{1}:i_{n_{l-1}})}^{\alpha,(l-1,1)} \frac{\partial\phi_{i_{m}}(x_{m})}{\partial x_{m}} \frac{1}{\phi_{i_{m}}(x_{m})} \prod_{n=1}^{n_{l-1}} \phi_{i_{n}}(x_{n})\right) \\
& \quad \times \left(\sum_{I(i_{n_{l-1}+1}:i_{n_l})} \Psi_{I(i_{n_{l-1}+1}:i_{n_l})}^{\beta,(l-1,2)} \prod_{n=n_{l-1}+1}^{n_l} \phi_{i_{n}}(x_{n})\right) \\
&= \sum_{\alpha,\beta} T_{\alpha,\beta}^{\gamma,(l,1)} J_{\alpha m}^{(l-1)}(x_{1},...,x_{n_{l-1}}) g_{\beta}^{(l-1)}(x_{n_{l-1}+1},...,x_{n_l}) \ .
\end{split}
\end{equation}
\endgroup

If $n_{l-1}< m \le n_l$, then
\begin{align}
    J&_{\gamma m}^{(l)}(\bm{x})\\
    &= \sum_{\alpha,\beta} T_{\alpha,\beta}^{\gamma,(l,1)}  g_{\alpha}^{(l-1)}(x_1,...,x_{n_{l-1}}) J_{\beta m}^{(l-1)}(x_{n_{l-1}+1},...,x_{n_l})
\end{align}
is obtained.
Hereafter, we proceed with calculations assuming $1 \le m \le n_{l-1}$.

First, taking the expectation with respect to the randomness of $T$, $\left\langle\cdot\right\rangle_{T} := \mathbb{E}_{p(T)}[\cdot]$:
\begingroup

\begin{equation}
\begin{split}
\left\langle J_{\gamma m}^{(l)}(x)\right\rangle_{T}
&= \sum_{\alpha,\beta} \left\langle T_{\alpha,\beta}^{\gamma,(l,1)} J_{\alpha m}^{(l-1)}(x_{1},...,x_{n_{l-1}}) g_{\beta}^{(l-1)}(x_{n_{l-1}+1},...,x_{n_l})\right\rangle_{T} \\
&= \sum_{\alpha,\beta} \left\langle T_{\alpha,\beta}^{\gamma,(l,1)}\right\rangle_{T} \left\langle J_{\alpha m}^{(l-1)}(x_{1},...,x_{n_{l-1}})\right\rangle_{T} \left\langle g_{\beta}^{(l-1)}(x_{n_{l-1}+1},...,x_{n_l})\right\rangle_{T} \\
&= 0
\end{split}
\end{equation}
\endgroup
Therefore, for any $l,\gamma,m$,
\begin{align} \label{eq:meanJacobi}
  \left\langle J_{\gamma m}^{(l)}(\bm{x})\right\rangle_{T} = 0
\end{align}
is obtained.

Next, we calculate the covariance $C_{J_m, \eta\theta}^{(l)} := \left\langle J_{\eta m}^{(l)} J_{\theta m}^{(l)} \right\rangle_T$.

\begin{equation}
\begin{split}
C_{J_m,\eta\theta}^{(l)}(\bm{x})
&= \left\langle J_{\eta m}^{(l)}(\bm{x})J_{\theta m}^{(l)}(\bm{x})\right\rangle_{T} \\
&= \left\langle \left(\sum_{\alpha,\beta}T_{\alpha,\beta}^{\eta,(l,1)}J_{\alpha m}^{(l-1)}{g_{\beta}^{(l-1)}}\right) \left(\sum_{\alpha^{\prime},\beta^{\prime}}T_{\alpha^{\prime},\beta^{\prime}}^{\theta,(l,1)}J_{\alpha^{\prime}m}^{(l-1)}{g_{\beta^{\prime}}^{(l-1)}}\right) \right\rangle_{T} \\
&= \sum_{\alpha,\beta} \left\langle T_{\alpha,\beta}^{\eta,(l,1)}T_{\alpha,\beta}^{\theta,(l,1)}\right\rangle_{T} {\left\langle (J_{\alpha m}^{(l-1)}(x_{1},...,x_{n_{l-1}}))^{2} \right\rangle_{T} \left\langle (g_{\beta}^{(l-1)}(x_{n_{l-1}+1},...,x_{n_l}))^{2} \right\rangle_{T}} \\
&= \sum_{\alpha,\beta} \frac{\sigma_T^2}{{d_\alpha d_\beta}} \delta_{\eta\theta} {C_{J_m,\alpha\alpha}^{(l-1)}(x_{1},...,x_{n_{l-1}}) V_{g,\beta\beta}^{(l-1)}(x_{n_{l-1}+1},...,x_{n_l})} \\
&= {\sigma_T^{2}\delta_{\eta\theta} V_{J_m}^{(l-1)}(x_{1},...,x_{n_{l-1}}) V_{g}^{(l-1)}(x_{n_{l-1}+1},...,x_{n_l})}
\end{split}
\end{equation}
In the final equality, we set
\begin{equation}
V_{J_m}^{(l-1)}(x_{1},...,x_{n_{l-1}}) = C_{J_m,\alpha\alpha}^{(l-1)}(x_{1},...,x_{n_{l-1}}), \ \forall \alpha
\end{equation}
due to the isotropy with respect to the bond degrees of freedom within the network.

The above calculation yields the recurrence relation
\begin{equation}
    V_{J_m}^{(l)}(x_{1},...,x_{n_l}) = \left(\frac{\sigma_T^2}{2}\right)^{n_{l-1}} V_{J_m}^{(l-1)}(x_{1},...,x_{n_{l-1}})
\end{equation}
for the diagonal components (the off-diagonal components are zero).

Here, we compute \(V_J^{(1)}(x_1, x_2)\). First, we have
\begin{equation}
\begin{split}
J_{\gamma m}^{(1)}&(x_1, x_2) \\
&= \sum_{i_1,i_2}
    T_{i_1,i_2}^{\gamma,(1,1)}
    \frac{\partial \phi_{i_m}(x_m)}{\partial x_m}
    \frac{1}{\phi_{i_m}(x_m)}
    \,\phi_{i_1}(x_1)\,\phi_{i_2}(x_2)\,.
\end{split}
\end{equation}
Hence, it can be calculated that
\begin{align}
V_{J_m}^{(1)}&(x_1, x_2)
= \sum_{i_1,i_2}
    \left\langle
      \bigl(T_{i_1,i_2}^{\gamma,(1,1)}\bigr)^2
    \right\rangle
    \left(
      \frac{\partial \phi_{i_m}(x_m)}{\partial x_m}
    \right)^{\!2}
    \left(
      \frac{1}{\phi_{i_m}(x_m)}
    \right)^{\!2}
    \bigl(\phi_{i_1}(x_1)\bigr)^{\!2}
    \bigl(\phi_{i_2}(x_2)\bigr)^{\!2}\ .
\end{align}

Here we set
\begin{equation}
\phi_{0}(x)=\cos \biggl(\frac{\pi x}{2}\biggr),
\qquad
\phi_{1}(x)=\sin\biggl(\frac{\pi x}{2}\biggr).
\end{equation}
It follows that
\begin{equation}
V_{J_m}^{(1)}(x_{1},x_{2})
=\frac{\sigma_T^{2}\,\pi^{2}}{16}.
\end{equation}
Hence \(V_{J_m}^{(l)}(x_{1},\dots,x_{n_l})\) is constant (independent of \(\bm{x}\) and $m$) and satisfies the recurrence relation
\begin{equation}
V_{J}^{(l)} = \left(\frac{\sigma_T^2}{2}\right)^{n_{l-1}} V_{J}^{(l-1)}.
\end{equation}
Solving this yields
\begin{equation}
V_{J}^{(l)}=\frac{\pi^{2}}{8}\,\Bigl(\frac{\sigma_T^{2}}{2}\Bigr)^{n_l -1 }.
\end{equation}
This is the variance of the Jacobian matrix of random tree tensor network. At the top layer of one tree, the expected Lipschitz constant is then $C=\frac{\tilde{N}_x}{4}\sqrt{V_J^{(\tilde{L})}}$.

\section{Hyperparameter optimization protocol}

\noindent
Model hyperparameters were optimized using Optuna with the default \texttt{TPESampler}. The search budget was fixed to $300$ trials for both TTN-RC and ESN, so the tuning effort is matched at this level.

For TTN-RC, the sampled hyperparameter vector is
\[
\theta_{\mathrm{TTN}}=(\chi,\sigma_T,\sigma_{\mathrm{in}},\lambda),
\]
where $\chi$ is the TTN bond dimension and $\sigma_T$ is the TTN tensor-initialization standard deviation.

For ESN, the sampled hyperparameter vector is
\[
\theta_{\mathrm{ESN}}=(\rho_{\mathrm{ESN}}, s, \sigma_{\mathrm{in}},\lambda),
\]
where $\rho_{\mathrm{ESN}}$ is the target spectral radius of the reservoir matrix, and $s$ controls the reservoir connection sparsity (larger value means more sparse).

In both models, $\sigma_{\mathrm{in}}$ is the input-weight standard deviation (input-to-reservoir scaling), and $\lambda$ is the ridge regression regularization coefficient used in readout training.

\noindent
Let $\mathcal{S}$ denote the set of candidate hyperparameter vectors sampled by Optuna, with $|\mathcal{S}|=300$.

For a trial $\theta$, validation RMSE was computed from $N_{\mathrm{seed}}=10$ independent random reservoir seeds:
\[
\bar{\mathrm{RMSE}}_{\mathrm{val}}(\theta)=\frac{1}{N_{\mathrm{seed}}}\sum_{s=1}^{N_{\mathrm{seed}}}\mathrm{RMSE}_{\mathrm{val}}^{(s)}(\theta).
\]
The selected parameters are
\[
\theta^\ast=\arg\min_{\theta\in\mathcal{S}}\bar{\mathrm{RMSE}}_{\mathrm{val}}(\theta).
\]
The reported test performance is then
\[
\bar{\mathrm{RMSE}}_{\mathrm{test}}=\frac{1}{N_{\mathrm{seed}}}\sum_{s=1}^{N_{\mathrm{seed}}}\mathrm{RMSE}_{\mathrm{test}}^{(s)}(\theta^\ast).
\]
Hence the tuning budget is equal for both methods in terms of Optuna trials (300) and also in terms of evaluated reservoir realizations ($300\times N_{\mathrm{seed}}=3000$ per model).
\begin{table*}[t]
\centering
\footnotesize
\setlength{\tabcolsep}{3pt}
\caption{Optuna search space used in this study (default TPESampler).\\Each trial samples one parameter set from the ranges below.}
\label{tab:optuna_search_space}
\begin{tabular}{l p{2.45cm} p{1.45cm} p{1.55cm} p{2.75cm} p{4.35cm}}
\hline
Model & Hyperparameter & Min & Max & Distribution & Meaning \\
\hline
TTN-RC & $\chi$ & 10 & 100 & \mbox{Integer Uniform} & TTN bond dimension \\
TTN-RC & $\sigma_{T}$ & 1.0 & 2.5 & \mbox{Linear Uniform} & TTN tensor standard deviation \\
TTN-RC & $\sigma_{\mathrm{in}}^{\mathrm{TTN}}$ & $1.0\times10^{-1}$ & $1.0\times10^{1}$ & \mbox{Log Uniform} & Input-weight standard deviation (TTN) \\
TTN-RC & $\lambda$ & $1.0\times10^{-6}$ & $1.0\times10^{-1}$ & \mbox{Log Uniform} & Ridge regularization coefficient \\
ESN & $\rho_{\mathrm{ESN}}$ & 0.7 & 4.3 & \mbox{Linear Uniform} & Reservoir spectral radius \\
ESN & $s$ & 0.5 & 0.99 & \mbox{Linear Uniform} & Reservoir sparsity (larger = more sparse) \\
ESN & $\sigma_{\mathrm{in}}^{\mathrm{ESN}}$ & $1.0\times10^{-3}$ & 2.0 & \mbox{Log Uniform} & Input-weight standard deviation (ESN) \\
ESN & $\lambda$ & $1.0\times10^{-6}$ & $1.0\times10^{-1}$ & \mbox{Log Uniform} & Ridge regularization coefficient \\
\hline
\end{tabular}
\setlength{\tabcolsep}{6pt}
\end{table*}

\ack
This work was partially supported by the New Energy and Industrial Technology Development Organization (NEDO) under Grant Number JPNP23003 and by the Japan Science and Technology Agency (JST) through the Advanced Technologies for Carbon-Neutrality (ALCA-Next) program, Grant Number JPMJAN24B1. D.S. was supported by JST BOOST, Japan, Grant Number JPMJBS2415. We are particularly grateful to Naoki Yamamoto for sharing his valuable insights into quantum information, physics, and machine learning. We also thank Shinji Sato, Taihei Kuroiwa, Ryosuke Koga, and Yuya Ito for fruitful technical discussions.

\bibliography{refs}

\end{document}